\documentclass[%
reprint,
superscriptaddress,
bibnotes,
longbibliography,
amsmath,amssymb,
aps,
prb,
floatfix,
]{revtex4-2}

\usepackage{tabularx}
\usepackage{booktabs}
\usepackage{verbatim}
\usepackage{textcomp,gensymb}
\usepackage{graphicx}% Include figure files
\usepackage{dcolumn}% Align table columns on decimal point
\usepackage{bm}% bold math
\usepackage[colorlinks,linkcolor=blue,anchorcolor=blue,citecolor=blue,urlcolor=blue,filecolor=blue,menucolor=blue,runcolor=blue]{hyperref}% add hypertext capabilities
\usepackage[mathlines]{lineno}% Enable numbering of text and display math
\usepackage{xcolor}
\usepackage{multirow}
\newcolumntype{Y}{>{\centering\arraybackslash}X}
\newcolumntype{s}{>{\hsize=.5\hsize\centering\arraybackslash}X}  % 小列宽
\newcolumntype{b}{>{\hsize=.5\hsize}X}  % 小列宽

\begin{document}
	
	\title{Nodeless multigap superconductivity in organic-ion-intercalated (tetrabutyl~ammonium)$_{0.3}$FeSe}% Force line breaks with \\
	
	\author{Jinyu Wu}
	\affiliation{Center for Correlated Matter and School of Physics, Zhejiang University, Hangzhou 310058, China.}
%\begin{comment}
	\author{Mengzhu Shi}
	\affiliation{Department of Physics, and CAS Key Laboratory of Strongly-coupled Quantum Matter Physics, University of Science and Technology of China, Hefei, Anhui 230026, China.}
	\author{Jianwei Shu}
	\affiliation{Center for Correlated Matter and School of Physics, Zhejiang University, Hangzhou 310058, China.}
	\author{Zhaoyang Shan}
	\affiliation{Center for Correlated Matter and School of Physics, Zhejiang University, Hangzhou 310058, China.}
	\author{Toni Shiroka}
	\affiliation{Laboratory for Solid State Physics, ETH Zürich, 8093 Zürich, Switzerland.}
	\affiliation{Laboratory for Muon-Spin Spectroscopy, Paul Scherrer Institut, Villigen PSI, 5232 Villigen, Switzerland.}
	\author{Devashibhai Adroja}
	\affiliation{ISIS Facility, Rutherford Appleton Laboratory, Chilton, Didcot Oxon, OX11 0QX, United Kingdom.}
	\affiliation{Highly Correlated Matter Research Group, Physics Department, University of Johannesburg, PO Box 524, Auckland Park 2006, South Africa.}
	\author{Xianhui Chen}
	\affiliation{Department of Physics, and CAS Key Laboratory of Strongly-coupled Quantum Matter Physics, University of Science and Technology of China, Hefei, Anhui 230026, China.}
	\affiliation{Collaborative Innovation Center of Advanced Microstructures, Nanjing 210093, China.}
	\author{Michael Smidman}
	\email{msmidman@zju.edu.cn}
	\affiliation{Center for Correlated Matter and School of Physics, Zhejiang University, Hangzhou 310058, China.}
%\end{comment}

	\date{\today}% It is always \today, today,
	
	\begin{abstract}
We probe the superconducting order parameter of the organic-ion-intercalated FeSe-based superconductor (tetrabutyl ammonium)$_{0.3}$FeSe [(TBA)$_{0.3}$FeSe] using  muon-spin relaxation/rotation ($\mu$SR). Zero-field $\mu$SR measurements show only a weak temperature dependence with no evidence for magnetic ordering or broken  time-reversal symmetry in the superconducting state. The temperature dependence of the superfluid density is deduced from transverse-field $\mu$SR measurements with fields applied both parallel and perpendicular to the $c$~axis axis, and can be well described by a nodeless two-gap $s+s$ wave model. These properties are reminiscent of those of (Li$_{1-x}$Fe$_x$)OHFe$_{1-y}$Se, which also has a comparably enhanced $T_c$, suggesting that such a gap structure is a common feature of quasi-two-dimensional intercalated FeSe-based superconductors.
		
	\end{abstract}
	
	\maketitle

	\section{Introduction}\label{intro}

	The discovery of superconductivity in LaFeAsO$_{1-x}$F$_x$ with $T_c=26\,\mbox{K}$ \cite{IronBasedLayeredSuperconductorkamihara2008} revealed the iron-pnictides to be the second family of high-temperature superconductors. Shortly after, superconductivity was found in iron-chalcogenide systems, the simplest example being FeSe with $T_c=8\,\mbox{K}$  \cite{SuperconductivityPbOtypeStructurehsu2008}. While an enhanced $T_c$ can be realized in FeSe by applying pressure \cite{PhysRevB.80.064506,ElectronicMagneticPhasemedvedev2009,CrystalElectronicStructurekumar2010a} or by producing it in monolayer films (with a $T_c$ above 100 K) \cite{InterfaceInducedHighTemperatureSuperconductivitywang2012,he2013PhaseDiagramElectronica,Ge2015}, the enhancement of $T_c$ in bulk systems can also be achieved by the intercalation of spacer layers between the layers of FeSe. The first such examples were  $A$Fe$_2$Se$_2$ ($A$ = Na, K, Rb, Cs and Tl) with $T_c$ up to $32\,\mbox{K}$ \cite{SuperconductivityIronSelenideguo2010,CoexistenceSuperconductivityAntiferromagnetismliu2011}, but studies of their intrinsic superconducting properties were hindered by inhomogeneities due to iron vacancies, leading to a phase separation between antiferromagnetic and superconducting regions \cite{Chen2011,Texier2012,Li2021}. Subsequently, higher transition temperatures were achieved via the intercalation of organic ions in Li$_{0.56}$(NH$_2$)$_{0.53}$(NH$_3$)$_{1.19}$Fe$_2$Se$_2$ with $T_c = 39 \,\mbox{K}$ \cite{AmmoniaRichHighTemperatureSuperconductingsedlmaier2014}  and Li$_{0.6}$(NH$_2$)$_{0.2}$(NH$_3$)$_{0.8}$Fe$_2$Se$_2$ with $T_c = 44 \,\mbox{K}$ \cite{EnhancementSuperconductingTransitionburrard-lucas2013}. In LiOH-intercalated (Li$_{1-x}$Fe$_x$)OHFe$_{1-y}$Se, bulk superconductivity with $T_c$ over 40~K  coexists with antiferromagnetic order \cite{SoftChemicalControlsun2015,Lu2015}, and its stability in air, together with a lack of phase separation, made it a promising candidate for examining the intrinsic properties of intercalated FeSe superconductors.

There has been considerable debate as to whether iron-chalcogenide superconductors have an analogous pairing state to the sign-changing $s_{\pm}$ state  generally attributed to the iron pnictides \cite{UnconventionalPairingOriginatingkuroki2008,UnconventionalSuperconductivitywithaSignReversalintheOrder,Mazin2011}. In bulk FeSe, some thermodynamic experiments suggest a nodal gap \cite{SuperconductivityenhancedNematicityGapwang2017,FieldinducedSuperconductingPhasekasahara2014,biswas2018EvidenceNodalGap}, however other measurements support a fully gapped behavior, but with a significant gap anisotropy and possible sample dependence \cite{khasanov2008EvidenceNodelessSuperconductivity,HighlyAnisotropicSuperconductingchen2017, SuperconductingGapStructurejiao2017a, DiscoveryOrbitalselectiveCoopersprau2017, SuperconductingGapAnisotropyhashimoto2018}. The nature of the pairing state has been of particular interest for single-layer FeSe/SrTiO$_3$ and  intercalated FeSe superconductors, since their Fermi surface lacks the hole pocket at the zone center present in bulk FeSe and iron pnictide systems \cite{Qian2011,ElectronicOriginHightemperatureliu2012,CommonElectronicOriginzhao,SurfaceElectronicStructureniu2015}. In single-layer FeSe/SrTiO$_3$, fully-gapped superconductivity was revealed \cite{SuperconductingGapAnisotropyzhang2016,InterfaceInducedHighTemperatureSuperconductivitywang2012}, with the largest gap being of the order of $20\,\mbox{meV}$. Meanwhile two-gap nodeless superconductivity was found in (Li$_{1-x}$Fe$_x$)OHFe$_{1-y}$Se from various probes \cite{ScrutinizingDoubleSuperconductingdu2016,SurfaceElectronicStructureyan2016,khasanov2016ProximityinducedSuperconductivityInsulating,Smidman2017}, but there are conflicting conclusions from quasiparticle interference measurements as to whether these correspond to a sign-changing superconducting pairing state \cite{Du2018}, or sign preserving $s$-wave superconductivity \cite{SurfaceElectronicStructureyan2016}. Consequently, it is important to probe the pairing states of other families of intercalated FeSe-based superconductors.

Single crystals of FeSe-based superconductors intercalated with tetrabutyl ammonium  [(TBA)$_{0.3}$FeSe] \cite{shi2018FeSebasedSuperconductorsSuperconducting} and  cetyltrimethyl ammonium [(CTA)$_{0.3}$FeSe] \cite{OrganicionintercalatedFeSebasedSuperconductorsshi2018}  ions have also been successfully synthesized, with high onset $T_c$ values of 50$\,\mbox{K}$ and  45$\,\mbox{K}$, respectively. Such an intercalation  leads to both a charge transfer from the  sizeable organic ions to the FeSe layers, and a greatly enhanced separation between FeSe layers of around 15.5\,\AA. The latter gives rise to a two-dimensional superconductivity, which is evidenced by the observation of a pseudogap phase above $T_c$ in nuclear magnetic resonance (NMR) and in Nernst effect measurements \cite{kang2020PreformedCooperPairs}, which suggest the presence of preformed Cooper pairs that lack long range phase coherence. However, detailed information about the superconducting gap structure of these superconductors is still lacking.

In this paper, we examine the superconducting pairing state of single crystalline (TBA)$_{0.3}$FeSe  using zero-field (ZF-) and transverse-field (TF-) muon-spin relaxation/rotation ($\mu$SR) measurements. From ZF-$\mu$SR measurements, there is no evidence of time-reversal symmetry breaking in the superconducting state, and no signatures of magnetic ordering are observed. The temperature dependence of the magnetic penetration depth is deduced from TF-$\mu$SR for $H\parallel c$ and $H\perp c$, which can be well described by a nodeless two-gap $s+s$-wave model along both directions.

	\section{\label{exp detail}Experimental details}
Single crystals of (TBA)$_{0.3}$FeSe  were prepared via an electrochemical intercalation method, as described in Ref.~\cite{shi2018FeSebasedSuperconductorsSuperconducting}. ZF- and TF-$\mu$SR experiments were performed using the general-purpose surface-muon (GPS) instrument at the Swiss Muon Source (S$\mu$S) of the Paul Scherrer Institut (PSI), Switzerland. Spin-polarized positive muons are implanted into the sample, and decay positrons are preferentially emitted  along the muon-spin direction. 
Therefore, the asymmetry of the emitted positrons depends on the dynamics and distribution of the local magnetic fields at the muon stopping site(s). Plate-like single crystals were mounted on a silver sample holder so that the $c$~axis axis was perpendicular to the plate, and parallel to the muon beam. Figure~\ref{fig_TF_c}(a) shows the configuration used for the TF--$\mu$SR measurements. These were performed with an applied field of 80~mT along the $c$~axis axis, for which the spin-polarization  direction of the muon beam was rotated by $45^{\circ}$, and an applied field of 10~mT in the  $ab$~plane, for which the muon spins were not rotated. Note that for the spin rotated case, the maximum asymmetry is approximately 70\% of the unrotated muon spin case [i.e., $A(0)\sin(45^{\circ})$]. The TF was applied above the superconducting transition temperature $T_c$, after which the samples were cooled to the base temperature of 1.6~K, and measurements were performed upon warming up the sample. All  the data were analyzed using the data analysis program WIMDA \cite{pratt2000WIMDAMuonData}.

\section{\label{Results and dis}Results}
\subsection{\label{ZF}Zero-field $\mu$SR}
\begin{figure}[tb]
		\includegraphics[width=0.51\textwidth]{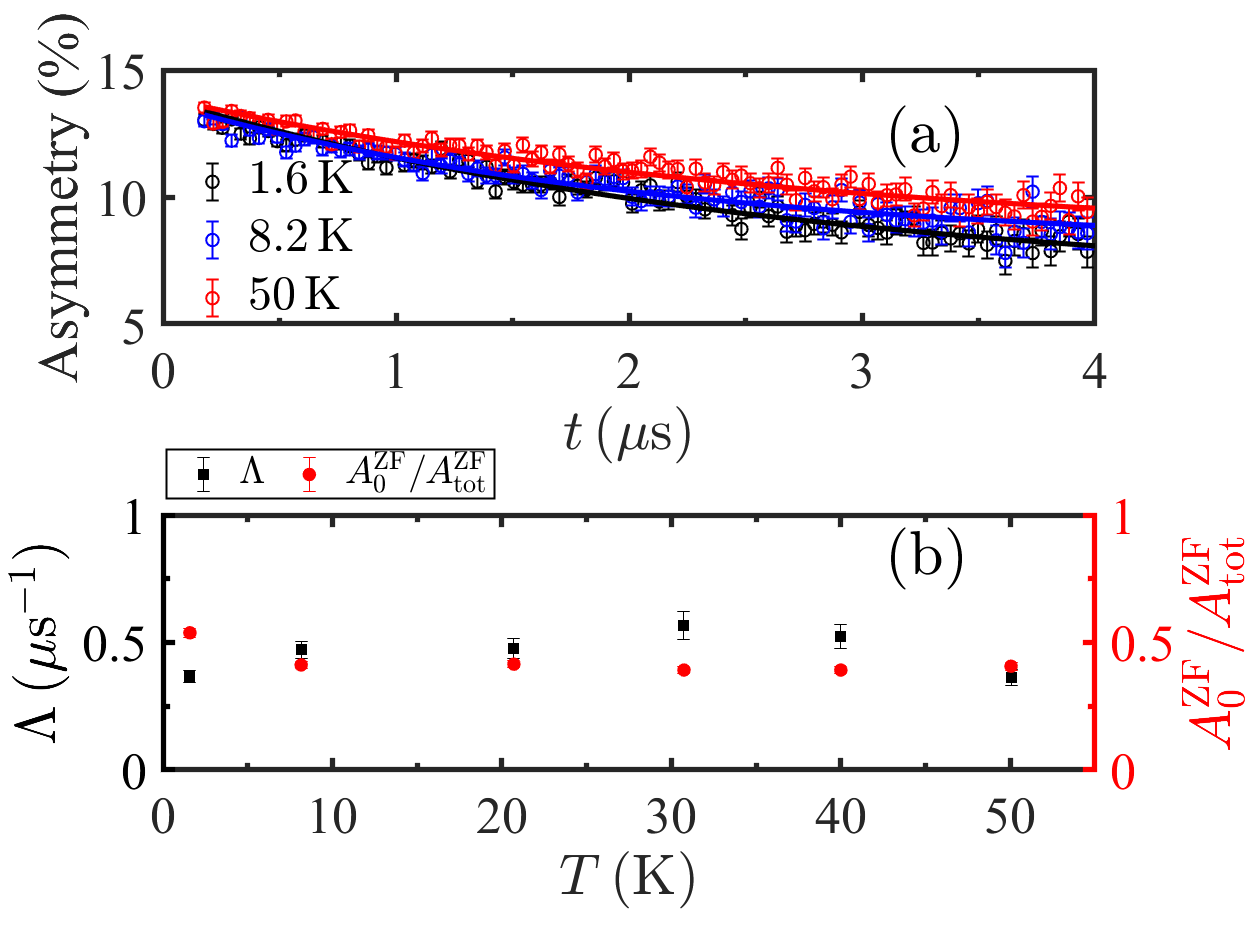}% Here is how to import EPS art
		\caption{\label{fig1} (a) Zero-field $\mu$SR spectra of (TBA)$_{0.3}$FeSe at $1.6$ (black), $8.2$ (blue) and $50\,\mbox{K}$ (red), with the solid lines showing the fitting results using Eq.~\ref{eq_AZF}. (b)  Temperature dependence of the fitted Lorentzian relaxation rate $\Lambda$ and the ratio $A_0^{\rm{ZF}}/A_{\rm{tot}}^{\rm{ZF}}$.}
	\end{figure}

In order to probe the magnetic properties and examine whether there is time-reversal symmetry breaking in the superconducting state, ZF-$\mu$SR measurements were performed on (TBA)$_{0.3}$FeSe. Figure~\ref{fig1}(a) displays ZF-$\mu$SR spectra collected at three temperatures, $1.6\,\mbox{K}$, $8.2\,\mbox{K}$ and $50\,\mbox{K}$ (below and above $T_c=48\,\mbox{K}$), with the initial spin-polarization of the muon beam rotated by $45^{\circ}$. Since the asymmetry decays relatively quickly at low times, the data were analyzed by an exponential decay function,
	\begin{equation}\label{eq_AZF}
		A(t) = A_{0}^{\rm{ZF}}\exp(-\Lambda t) + A_{\rm{bg}}^{\rm{ZF}}.
	\end{equation}
	Here $A_{0}^{\rm{ZF}}$ is the initial asymmetry, corresponding to muons stopping in the sample, $A_{\rm{bg}}^{\rm{ZF}}$ is a time-independent contribution from muons stopping in the silver sample holder, and $\Lambda$ is the Lorentzian relaxation rate. Such an exponential decay can correspond to either magnetic fields fluctuating much faster than the muon precession rate, or to dilute static fields with a Lorentzian distribution.  The temperature dependence of $\Lambda(T)$ and the ratio $A_0^{\rm{ZF}}/A_{\rm{tot}}^{\rm{ZF}}$ are shown in Fig.~\ref{fig1}(b), where $A_{\rm{tot}}^{\rm{ZF}}$ is the total initial asymmetry. $\Lambda(T)$ exhibits a relatively weak temperature dependence without a pronounced increase upon crossing $T_c$. Clearly, there is no evidence for either time-reversal symmetry breaking or significant slowing down of spin-fluctuations. In addition, the data points between 20 and 40 K were measured with the  spin-polarization of the muon beam parallel to the $c$~axis axis (unrotated). The very similar behaviors to the measurements with the muon spins rotated also suggests a lack of significant anisotropy of the local fields.
	
\subsection{\label{TF_c}Transverse-field $\mu$SR with $H\parallel c$}

\begin{figure}
		\includegraphics[width=0.5\textwidth]{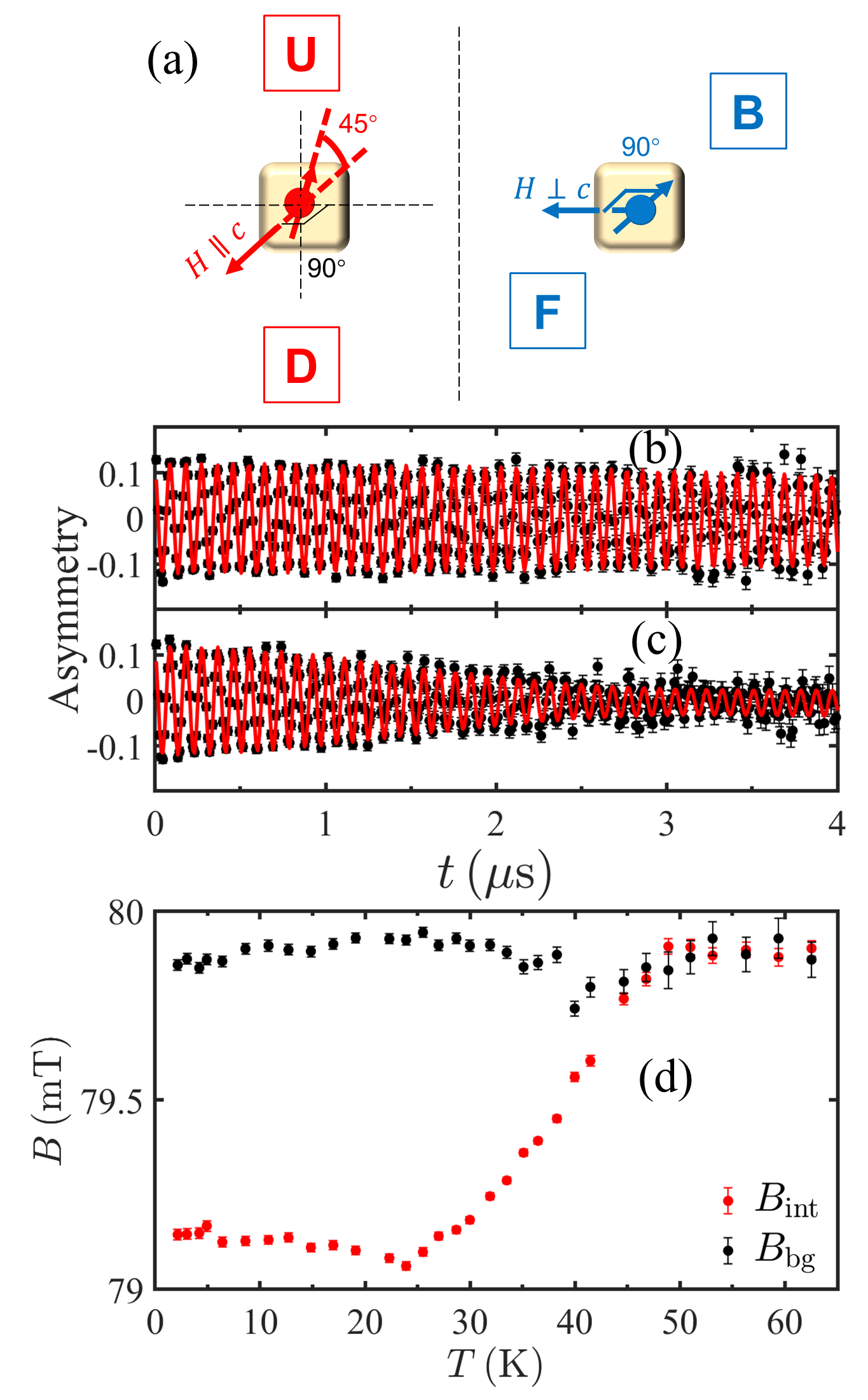}% Here is how to import EPS art
		\caption{\label{fig_TF_c} (a) Illustration of the geometry for  TF-$\mu$SR measurements, where for  $H\parallel c$ the muon spins are rotated upwards by 45$^\circ$, and the up(U)-down(D) detector pair is used to determine the asymmetry, while for $H\perp c$ the muon spins are unrotated and the asymmetry is obtained from the forward(F)-backward (B) pair. Transverse-field $\mu$SR spectra are collected in a transverse field of $80\,\mbox{mT}$ applied parallel to the $c$~axis axis at (b) $62.5$ and (c) $2.2\,\mbox{K}$. (d) Temperature dependence of the internal magnetic field inside the sample $B_{\rm{int}}$ and the magnetic field in the silver holder (background) $B_{\rm{bg}}$ as obtained from the fits.}
	\end{figure}

In order to probe the superconducting gap structure of (TBA)$_{0.3}$FeSe, TF-$\mu$SR measurements  were performed with a TF applied both parallel and perpendicular to the $c$~axis axis, with the configurations shown in Fig.~\ref{fig_TF_c}(a). These were performed upon field-cooling the sample below $T_c$, in order to induce a regular flux-line, from which the magnetic penetration depth $\lambda(T)$ can be extracted. 

Figures~\ref{fig_TF_c}(b) and (c) display the $\mu$SR time spectra with an applied TF of $\mu_0H=80\,\mbox{mT}$ parallel to the $c$~axis~axis above  ($62.5\,\mbox{K}$) and below ($2.2\,\mbox{K}$)  $T_c$, respectively. At $62.5\,\mbox{K}$, (TBA)$_{0.3}$FeSe is in the normal state and the muons precess at a single frequency with a small depolarization due to  quasistatic nuclear moments. While at $2.2\,\mbox{K}$, the more rapid damping arises from the nonuniform field distribution due to the flux line lattice. The data were analyzed using

\begin{equation}\label{eq_TF}
A(t) = A_0e^{-(\sigma t)^2/2}\cos(\gamma_{\mu}B_{\rm{int}}t+\phi) + A_{\rm{bg}}\cos(\gamma_{\mu}B_{\rm{bg}}t+\phi),
\end{equation}

\noindent where $\sigma$ is the Gaussian relaxation rate corresponding to the sample, $ A_0$ and $ A_{\rm{bg}}$ are the initial asymmetries corresponding to the sample and background components, $B_{\rm{int}}$ and $B_{\rm{bg}}$ are the corresponding internal and background magnetic fields,  $\phi$ is the initial phase of the precession signal, and $\gamma_{\mu}/2\pi = 135.5\, \mbox{MHz/T}$ is the muon gyromagnetic ratio. The total initial asymmetry $A_{\rm{tot}} = A_0+A_{\rm{bg}}$ and $\phi$ were fixed from fitting the data at the highest temperatures (above $T_c$) while $ A_0/A_{\rm{bg}}$ was fixed to the value obtained from fitting the data at the lowest temperature, where the background is better visible at long times. It can be seen from the red solid lines in  Fig.~\ref{fig_TF_c} that Eq.~\ref{eq_TF} well describes the TF-data above and below $T_c$. The temperature dependence of $B_{\rm{int}}$ and $B_{\rm{bg}}$ is shown in Fig.~\ref{fig_TF_c}(d). At temperatures above $T_c$, $B_{\rm{int}}$  and $B_{\rm{bg}}$ show similar values, slightly smaller than the applied field, indicating a weak diamagnetism that is consistent with muons stopping in the organic-ion  layers. Below $T_c$ there is a decrease of $B_{\rm{int}}$  while  $B_{\rm{bg}}$ remains almost unchanged, indicating a diamagnetic response due to the formation of a flux line lattice, as expected for a  type-\uppercase\expandafter{\romannumeral2} superconductor.

The superconducting contribution to the Gaussian relaxation rate is calculated using $\sigma_{sc}^2 = \sigma^2 - \sigma_{n}^2$, where $\sigma_{n}^2$ is the nuclear dipolar contribution, estimated from fitting the data above $T_c$, where $\sigma_{n}=0.200(7) \,\mu\mbox{s$^{-1}$}$ for $H\parallel c$. If the applied field is much less than the upper critical field ($H\ll H_{c2}$), $\sigma_{sc}$ is proportional to the inverse square of the penetration depth  \cite{brandt2003PropertiesIdealGinzburgLandau}. When the TF is applied parallel to the $c$~axis axis, $\sigma_{sc}^{\parallel c}$ probes the in-plane penetration depth $\lambda_{ab}$ as  \cite{khasanov2016ProximityinducedSuperconductivityInsulating}

	\begin{equation} \label{Lambda_ab_inplane}
		\lambda_{ab}^{-2} = 9.32(\rm{\mu m^{-2}/\mu s^{-1}})\times \sigma_{sc}^{\parallel c}(\rm{\mu s^{-1}}).
	\end{equation}

	\begin{figure}[tb]
		\includegraphics[width=0.5\textwidth]{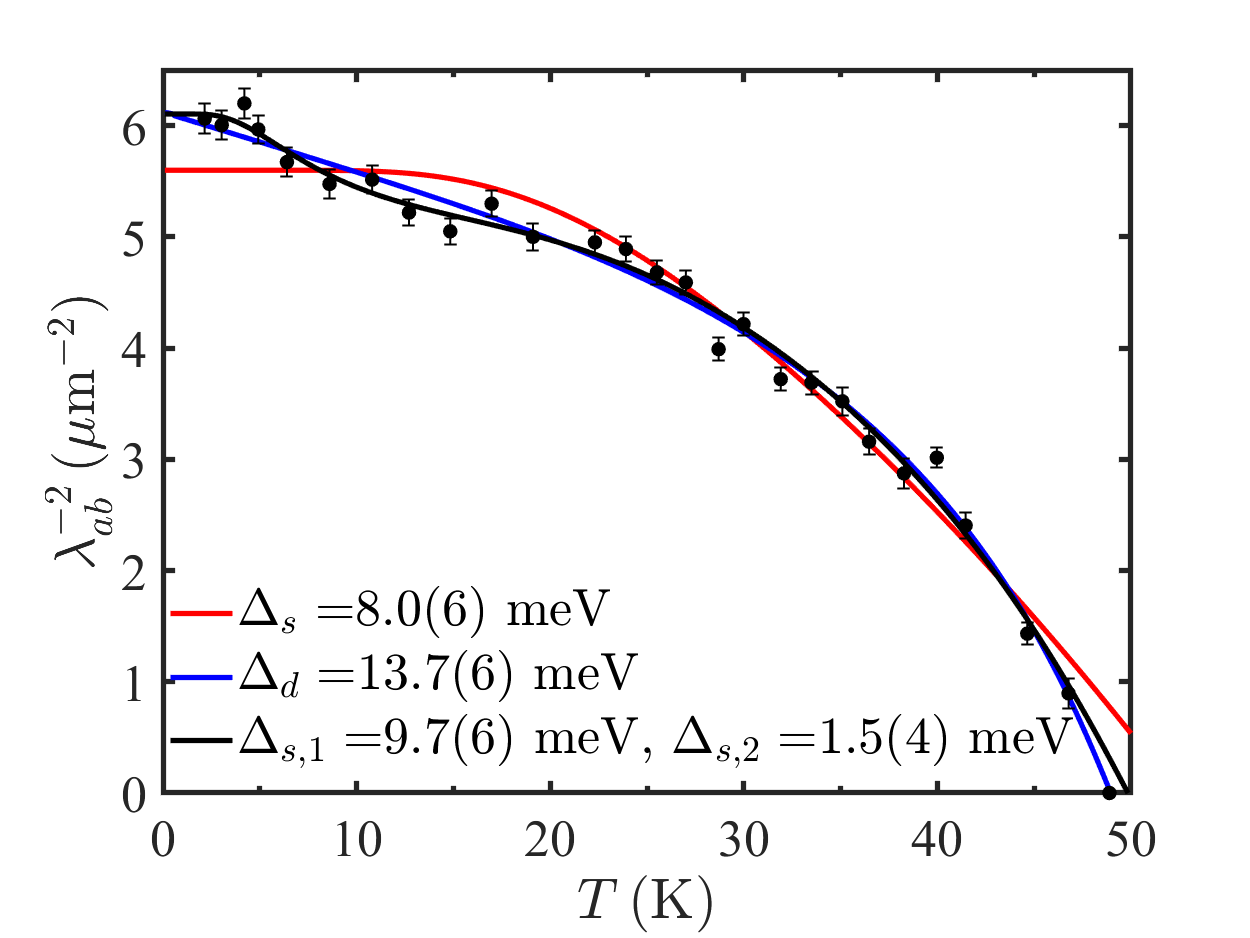}% Here is how to import EPS art
		\caption{\label{fig_TF_c_gap_fit_no_chi} Temperature dependence of $\lambda_{ab}^{-2}(T)$ of (TBA)$_{0.3}$FeSe obtained from a $80\,\mbox{mT}$ transverse field parallel to the $c$~axis axis. The solid lines are the fitted curves corresponding to a single gap $s$-wave model (red), a single gap $d$-wave model (blue) and a two-gap $s+s$ wave model (black).}
	\end{figure}

	The temperature dependence of the inverse square of the penetration depth $\lambda_{ab}^{-2}(T)$ is shown in Fig.~\ref{fig_TF_c_gap_fit_no_chi}. At low temperatures, $\lambda_{ab}^{-2}(T)$ saturates below $\sim4.9\,\mbox{K}$, which is consistent with a nodeless gap structure, since thermal excitations are unable to deplete the superconducting condensate in a fully gapped superconductor at sufficiently low temperatures. If there are nodes in the gap, $\lambda_{ab}^{-2}$ would continuously increase upon lowering the temperature due to the presence of low energy excitations. Furthermore, at intermediate temperatures, around $12\,\mbox{K}$, $\lambda_{ab}^{-2}(T)$ exhibits an inflection point, indicative of multigap superconductivity.
	
	The normalized superfluid density can be expressed as $\tilde{n}(T) = \lambda^{-2}(T)/\lambda^{-2}(0)$  and  can be analyzed utilizing the London approach \cite{tinkham2004introduction}:
	\begin{equation}\label{superfluid}
		\tilde{n}(T) = 1 + \frac{1}{\pi} \int_0^{2\pi}\int_{\Delta(T,\phi)}^{\infty} \frac{\partial f}{\partial E} \frac{E\mbox{d}E\mbox{d}\phi}{\sqrt{E^2-\Delta^2(T,\phi)}}
	\end{equation}
\noindent Here $\lambda^{-2}(0)$ is the magnetic penetration depth at zero temperature, and $f=[1 + \exp(-E/k_BT)]^{-1}$ is the Fermi-Dirac function. The superconducting gap function can be written as $\Delta(T,\phi)=\Delta(T)g(\phi)$, where the temperature dependent part is given by  $\Delta(T) = \Delta(0)\tanh\left\lbrace 1.82[1.018(T_c/T-1)]^{0.51}\right\rbrace$, where $\Delta(0)$ is the zero temperature gap. The angular dependence $g(\phi)$ ($\phi$ is the azimuthal angle) is $g(\phi) = 1$ for an $s$-wave gap ($\Delta_s$) and $g(\phi) = \cos(2\phi)$ for a $d$-wave gap ($\Delta_d$).

Figure~\ref{fig_TF_c_gap_fit_no_chi} displays the results of fitting for the superconducting gap structure with different models using Eq.~\ref{superfluid}. Both the nodeless single gap $s$-wave model and the $d$-wave model with line nodes do not adequately describe the data, where the former exhibits a large deviation over a wide temperature range, while the latter cannot account for the inflection point nor the low temperature saturation. To account for these features, $\lambda_{ab}^{-2}(T)$ was analyzed with a phenomenological  two-gap $s+s$ wave model:
	\begin{equation}\label{twogapsuperfluid}
		\tilde{n}(T) = x\tilde{n}_1(T, \Delta_{s,1}) + (1-x)\tilde{n}_2(T, \Delta_{s,2}),
	\end{equation}	
\noindent where $\tilde{n}_i(T, \Delta_i)$ is the superfluid density component corresponding to the gap $\Delta_{s,i}$ calculated using Eq.~\ref{superfluid}, with a weight $x$ ($0\leq x\leq 1$) for $i=1$ and $1-x$ for $i=2$. As shown in Fig~\ref{fig_TF_c_gap_fit_no_chi}, the two-gap $s+s$ model can well fit $\lambda_{ab}^{-2}(T)$, with  fitted parameters $\lambda_{ab}(0)=405(5)\,\mbox{nm}$, $\Delta_{s,1}(0)=9.7(6)\,\mbox{meV}$, $\Delta_{s,2}(0)= 1.5(4)\,\mbox{meV}$, $x=0.81(3)$, and $T_c=49.8(8)\,\mbox{K}$. The fitted parameters for other models are given in Table~\ref{tab:my-table}.

	\subsection{\label{TF}Transverse-field $\mu$SR with $H\perp c$}
	
\begin{figure}[tb]
		\includegraphics[width=0.5\textwidth]{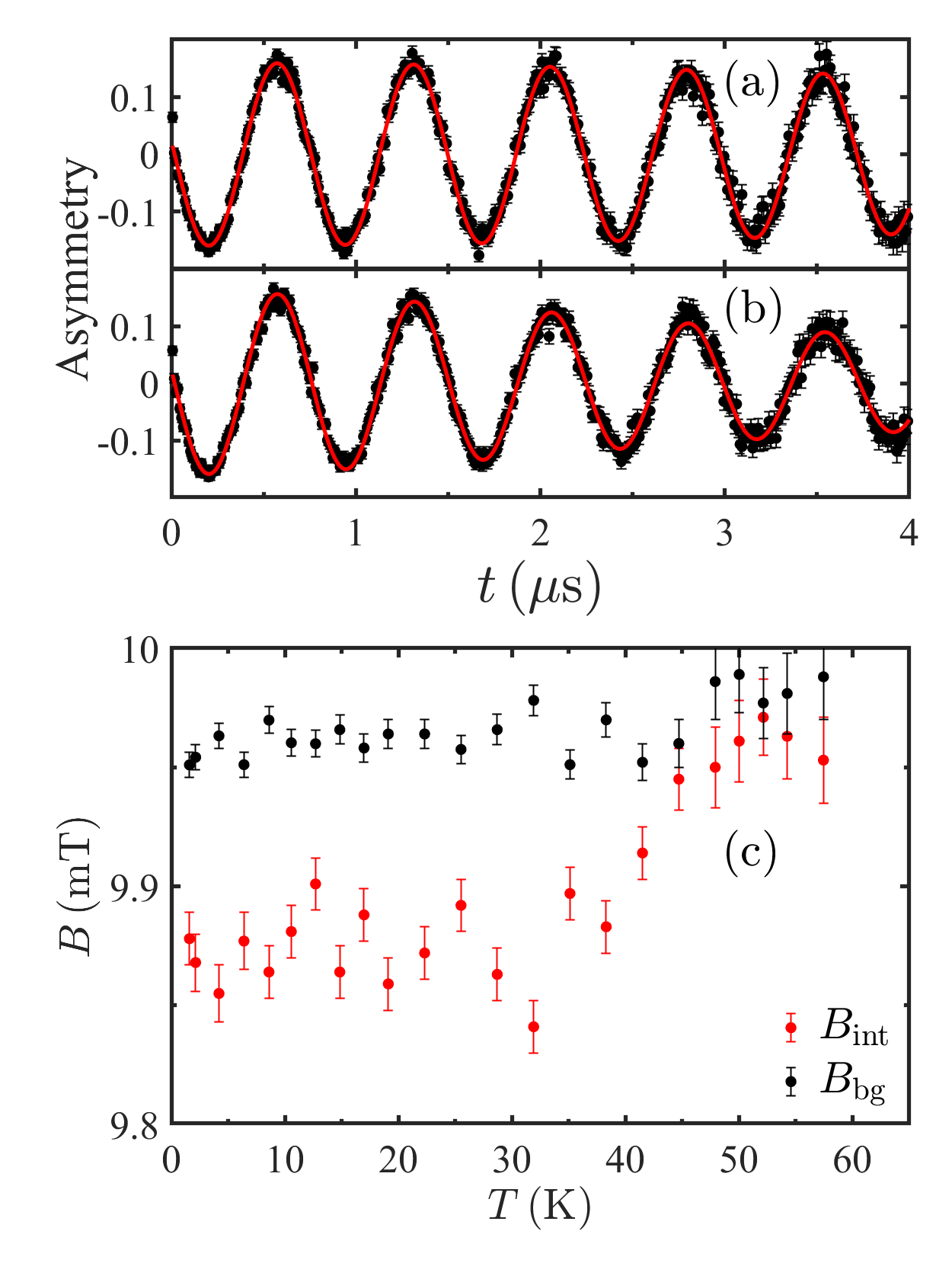}% Here is how to import EPS art
		\caption{\label{fig_TF_ab} Transverse-field $\mu$SR spectra in a transverse field of $10\,\mbox{mT}$ parallel to the $ab$~plane at (a) $57.4$ and (b) $1.6\,\mbox{K}$. The solid lines are the fitted curves using Eq.~\ref{eq_TF}. (c) Temperature dependence of the internal magnetic field $B_{\rm{int}}$ and the background magnetic field (silver holder) $B_{\rm{bg}}$ obtained from the fitting.}
	\end{figure}

	Figures \ref{fig_TF_ab}(a) and (b) display the $\mu$SR spectra with a TF field of $10\,\mbox{mT}$ perpendicular to the $c$~axis axis at $57.4\,\mbox{K}$ and $1.6\,\mbox{K}$, respectively. The precession at a single frequency above $T_c$ at $57.4\,\mbox{K}$ and the depolarization at $1.6\,\mbox{K}$ correspond to type-\uppercase\expandafter{\romannumeral2} superconductivity. The red solid lines show the fitting using Eq.~\ref{eq_TF}, and $\sigma_{n}$ is estimated to be $0.195(5)\,\mu\mbox{s$^{-1}$}$ from the normal state analysis. The temperature evolution of $B_{\rm{int}}$ and $B_{\rm{bg}}$ shown in Fig.~\ref{fig_TF_ab}(c) are also consistent with  type-\uppercase\expandafter{\romannumeral2} superconductivity and the system being in the mixed state.
	
\begin{figure}[h]
		\includegraphics[width=0.5\textwidth]{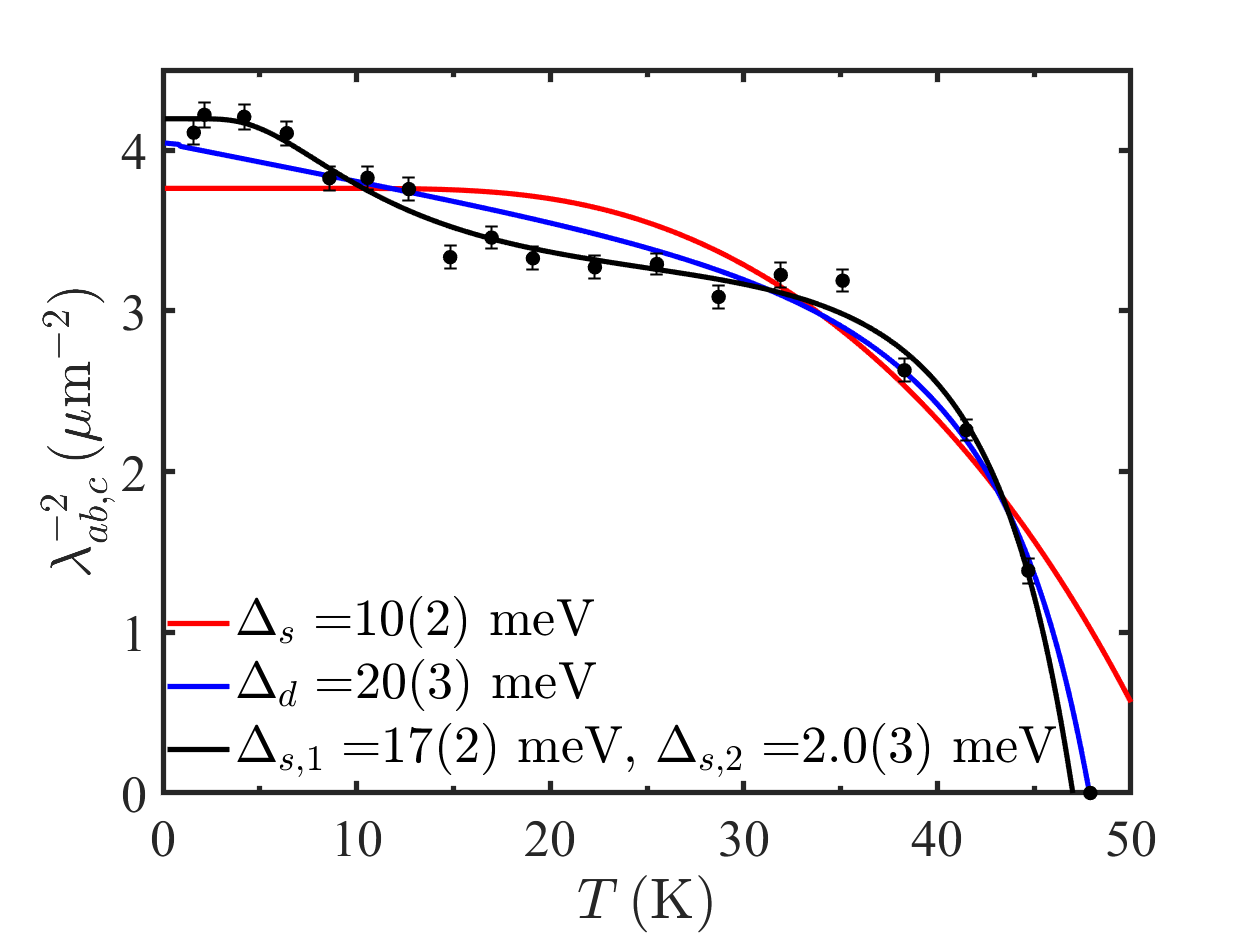}% Here is how to import EPS art
		\caption{\label{fig_TF_ab_gap_fit_no_chi} Temperature dependence of $\lambda_{ab,c}^{-2}$ of (TBA)$_{0.3}$FeSe with a transverse field of $10\,\mbox{mT}$ parallel to the $ab$~plane. The solid lines are the fitting curves of a single $s$-wave model (red line), a single $d$-wave model (blue line) and a two-gap $s+s$ wave model (black line), respectively.}
	\end{figure}

	The temperature dependence of  $\lambda_{ab,c}^{-2}(T)$ is displayed in Fig.~\ref{fig_TF_ab_gap_fit_no_chi}. It shows a similar behavior to $\lambda_{ab}^{-2}(T)$, including a saturation behavior below about $6.4\,\mbox{K}$ and a plateau with an inflection point at intermediate temperatures, which also indicate a nodeless multigap structure. Correspondingly, the single gap $s$-wave and $d$-wave models also poorly describe the data, as shown in Fig.~\ref{fig_TF_ab_gap_fit_no_chi}. $\lambda_{ab,c}^{-2}(T)$  were fitted using a two-gap $s+s$ wave model (Eq.~\ref{twogapsuperfluid}), which well fits the data with $\lambda_{ab,c}(0) = 488(4)\,\mbox{nm}$, $\Delta_{s,1}(0) = 17(2)\,\mbox{meV}$, $\Delta_{s,2}(0)= 2.0(3)\,\mbox{meV}$, $x = 0.74(2)$, and $T_c = 47.0(6)\,\mbox{K}$. The fitted parameters for other models are also summarized in Table~\ref{tab:my-table}.

	\begin{table*}[htp]
		\centering
		\caption{Summary of the fit parameters of the superfluid density for (TBA)$_{0.3}$FeSe with $H\parallel c$ ($\lambda_{ab}^{-2}$) and $H\perp c$ ($\lambda_{ab,c}^{-2}$), including the goodness of fit $\chi^2$. }
		\label{tab:my-table}
		\renewcommand\arraystretch{1.25}
		\begin{tabularx}{\textwidth}{sYYYsYss}
				\toprule[1pt]
				\toprule
				&Model&Gap value(s)&Gap value(s)&$T_c$&$\lambda(0)$&$x$& $\chi^2$\\
				& &(meV)&($k_BT_c$)&(K)&(nm)& &\\%
				\midrule
				\multirow{3}{*}{$\lambda_{ab}^{-2}$}&$s${-}wave gap&8.0(6)&1.8(1)&52(2)&423(4)&-&6.17\\%
				&$d${-}wave gap&13.7(6)&3.2(1)&49.0(6)&404(2)&-&2.38\\%
				&$s+s${-}wave gap&9.7(6)/1.5(4)&2.3(2)/0.3(1)&49.8(8)&405(5)&0.81(3)&2.40\\%
				\hline
				\multirow{3}{*}{$\lambda_{ab,c}^{-2}$}&$s${-}wave gap&10(2)&2.3(5)&52(5)&516(7)&-&19.1\\%
				&$d${-}wave gap&20(3)&4.9(6)&48(1)&497(5)&-&7.35\\%
				&$s+s${-}wave gap&17(2)/2.0(3)&4.2(5)/0.49(8)&47.0(6)&488(4)&0.74(2)&2.56\\%
				\bottomrule
				\bottomrule
		\end{tabularx}
	\end{table*}

	\section{\label{Con}  Discussion and Conclusions}
	
	Our TF-$\mu$SR measurements for fields applied both parallel and perpendicular to the $c$~axis axis evidence a nodeless two-gap order parameter in (TBA)$_{0.3}$FeSe.  The magnitudes of the larger of the two gaps obtained for the two field directions are $4.2(5)k_BT_c$ ($17(2)\,\mbox{meV}$) and $2.3(2)k_BT_c$ ($9.7(6)\,\mbox{meV}$) for $\lambda_{ab,c}^{-2}$ and $\lambda_{ab}^{-2}$, respectively, where the former is very close to the gap magnitude of $16\,\mbox{meV}$ from scanning tunnelling spectroscopy  (STS) \cite{kang2020PreformedCooperPairs}. On the other hand, the smaller gaps are  $0.49(8)k_BT_c$ ($2.0(3)\,\mbox{meV}$) and $0.3(1)k_BT_c$ ($1.5(4)\,\mbox{meV}$) for the respective directions, which are significantly smaller than the large gaps. Note that although the $\chi^2$ of the $d${-}wave and $s+s${-}wave model fitting are similar for $H\parallel c$ (Table~\ref{tab:my-table}), it can be seen that the $d$-wave model systematically gives a poorer fit at low temperatures, failing to capture the inflection and low temperature flattening of the data. Moreover, for $H\perp c$ the $\chi^2$ of the $s+s${-}wave fit is considerably lower.

The occurrence of two-gap nodeless superconductivity is in contrast to a TF-$\mu$SR study of FeSe, where there is a lack of saturation and a linear temperature dependence of $\lambda_{ab}^{-2}(T)$, which was best described by a model with one nodal and one nodeless gap \cite{biswas2018EvidenceNodalGap}, although the presence of deep gap minima has also been suggested  \cite{HighlyAnisotropicSuperconductingchen2017, SuperconductingGapStructurejiao2017a, DiscoveryOrbitalselectiveCoopersprau2017}. The nodal order parameter was proposed to correspond to an $s+d$ wave pairing state with nodes on the electron pocket \cite{biswas2018EvidenceNodalGap}. The presence of nodes has also been inferred from STS measurements of FeSe thin-films \cite{science1202226}, while such measurements on monolayer FeSe/SrTiO$_3$ consistently show  U-shaped spectra \cite{InterfaceInducedHighTemperatureSuperconductivitywang2012,fan2015plain} indicating nodeless pairing upon reducing the dimensionality.

Two-gap nodeless superconductivity has also been reported from penetration depth studies of the  intercalated FeSe based superconductor (Li$_{0.84}$Fe$_{0.16}$)OHFe$_{0.98}$Se \cite{khasanov2016ProximityinducedSuperconductivityInsulating,Smidman2017}, which also has an  enhanced transition temperature of $T_c = 42\,\mbox{K}$. Similarly, nodeless superconductivity was inferred from angle-resolved photoemission spectroscopy measurements of the intercalated systems (Tl, K)Fe$_{1.78}$Se$_2$ \cite{StrongNodelessPairingwang2011} and A$_x$Fe$_2$Se$_2$ (A = K, Cs) \cite{NodelessSuperconductingGapzhang2011}. For  (Li$_{0.84}$Fe$_{0.16}$)OHFe$_{0.98}$Se, the  two gap magnitudes deduced from the analysis of  $\lambda_{ab}^{-2}$ are more similar with $\Delta_2/\Delta_1\approx0.5-0.6$ \cite{khasanov2016ProximityinducedSuperconductivityInsulating,Smidman2017}, as opposed to a corresponding ratio of 0.15 from our study of (TBA)$_{0.3}$FeSe. On the other hand, a small gap of 1.05~meV is deduced from the analysis of the out-of-plane penetration depth of(Li$_{0.84}$Fe$_{0.16}$)OHFe$_{0.98}$Se \cite{khasanov2016ProximityinducedSuperconductivityInsulating}, as compared to a larger gap of 9.7 meV. Interestingly these gap parameters are close to those obtained here from $\lambda_{ab}^{-2}$  of (TBA)$_{0.3}$FeSe. These similarities between the studies of Li$_{0.84}$Fe$_{0.16}$)OHFe$_{0.98}$Se and (TBA)$_{0.3}$FeSe suggest that such multigap nodeless structures are a common feature of  quasi-two-dimensional FeSe-based superconductors, seemingly different from bulk FeSe.

In (Li$_{0.84}$Fe$_{0.16}$)OHFe$_{0.98}$Se, the opening of a small gap only for the out-of-plane direction is ascribed to superconductivity in the spacer layers being induced by proximity to the FeSe layers \cite{khasanov2016ProximityinducedSuperconductivityInsulating}. A two-dimensional nature of the superconductivity in (TBA)$_{0.3}$FeSe is inferred from the observation of a Berezinskii-Kosterlitz-Thouless  transition, as well as a pseudogap phase above the superconducting $T_c$, where strong phase fluctuations prevent zero-resistance \cite{kang2020PreformedCooperPairs}, and the resistivity ratio $\rho_c/\rho_{ab}$ of around $10^5$ 
exceeds that of $\simeq2500$ for (Li$_{0.84}$Fe$_{0.16}$)OHFe$_{0.98}$Se \cite{PhysRevB.92.064515}. However, in (TBA)$_{0.3}$FeSe this two-dimensionality is not as readily inferred from the anisotropy of the superfluid density. Another notable feature of (Li$_{0.84}$Fe$_{0.16}$)OHFe$_{0.98}$Se is that clear evidence for magnetic ordering is found from  the observation of coherent oscillations in ZF-$\mu$SR  \cite{khasanov2016ProximityinducedSuperconductivityInsulating}. No such oscillations are found in our ZF-$\mu$SR  measurements of (TBA)$_{0.3}$FeSe, and there is only a weak temperature dependence of $\Lambda(T)$ [(Fig.~\ref{fig1}(b)]. Note that muons stopping in the large organic space layers may not be sensitive to the magnetic fields generated in the FeSe layers, and therefore the ZF-$\mu$SR  results may not reflect the intrinsic spin dynamics of the iron ions.

In summary, we performed  $\mu$SR measurements to probe the superconducting order parameter of (TBA)$_{0.3}$FeSe. No evidence is found for time-reversal symmetry breaking below $T_c$ from ZF-$\mu$SR, while the analysis of the temperature dependence of the superfluid density deduced from  TF-$\mu$SR with magnetic fields $H\parallel c$ and $H\perp c$ shows that there are multiple nodeless superconducting gaps, which is in line with the findings for the intercalated FeSe based superconductor (Li$_{0.84}$Fe$_{0.16}$)OHFe$_{0.98}$Se. Given that there are conflicting reports as to whether (Li$_{0.84}$Fe$_{0.16}$)OHFe$_{0.98}$Se exhibits a sign-changing \cite{Du2018,Davies2016,Pan2017}, or sign preserving  \cite{SurfaceElectronicStructureyan2016} $s$-wave pairing state, it is of particular interest to examine this aspect of the order parameter of (TBA)$_{0.3}$FeSe, such  as by utilizing quasiparticle interference techniques or by looking for a spin-resonance mode.

\section{Acknowledgments}
This work was supported by  the Zhejiang Provincial Natural Science Foundation of China (Grant No. LR22A040002), the National Key R\&D Program of China (Grant No. 2022YFA1402200 and  No. 2023YFA1406303), the Key R\&D Program of Zhejiang Province, China (Grant No. 2021C01002), and the National Natural Science Foundation of China (Grants No. 12222410). This work is based on experiments performed at the Swiss Muon Source S$\mu$S, Paul Scherrer Institute, Villigen, Switzerland.
	
%\bibliography{TBA}

\begin{thebibliography}{51}%
	\makeatletter
	\providecommand \@ifxundefined [1]{%
		\@ifx{#1\undefined}
	}%
	\providecommand \@ifnum [1]{%
		\ifnum #1\expandafter \@firstoftwo
		\else \expandafter \@secondoftwo
		\fi
	}%
	\providecommand \@ifx [1]{%
		\ifx #1\expandafter \@firstoftwo
		\else \expandafter \@secondoftwo
		\fi
	}%
	\providecommand \natexlab [1]{#1}%
	\providecommand \enquote  [1]{``#1''}%
	\providecommand \bibnamefont  [1]{#1}%
	\providecommand \bibfnamefont [1]{#1}%
	\providecommand \citenamefont [1]{#1}%
	\providecommand \href@noop [0]{\@secondoftwo}%
	\providecommand \href [0]{\begingroup \@sanitize@url \@href}%
	\providecommand \@href[1]{\@@startlink{#1}\@@href}%
	\providecommand \@@href[1]{\endgroup#1\@@endlink}%
	\providecommand \@sanitize@url [0]{\catcode `\\12\catcode `\$12\catcode
		`\&12\catcode `\#12\catcode `\^12\catcode `\_12\catcode `\%12\relax}%
	\providecommand \@@startlink[1]{}%
	\providecommand \@@endlink[0]{}%
	\providecommand \url  [0]{\begingroup\@sanitize@url \@url }%
	\providecommand \@url [1]{\endgroup\@href {#1}{\urlprefix }}%
	\providecommand \urlprefix  [0]{URL }%
	\providecommand \Eprint [0]{\href }%
	\providecommand \doibase [0]{https://doi.org/}%
	\providecommand \selectlanguage [0]{\@gobble}%
	\providecommand \bibinfo  [0]{\@secondoftwo}%
	\providecommand \bibfield  [0]{\@secondoftwo}%
	\providecommand \translation [1]{[#1]}%
	\providecommand \BibitemOpen [0]{}%
	\providecommand \bibitemStop [0]{}%
	\providecommand \bibitemNoStop [0]{.\EOS\space}%
	\providecommand \EOS [0]{\spacefactor3000\relax}%
	\providecommand \BibitemShut  [1]{\csname bibitem#1\endcsname}%
	\let\auto@bib@innerbib\@empty
	%</preamble>
	\bibitem [{\citenamefont {Kamihara}\ \emph {et~al.}(2008)\citenamefont
		{Kamihara}, \citenamefont {Watanabe}, \citenamefont {Hirano},\ and\
		\citenamefont {Hosono}}]{IronBasedLayeredSuperconductorkamihara2008}%
	\BibitemOpen
	\bibfield  {author} {\bibinfo {author} {\bibfnamefont {Y.}~\bibnamefont
			{Kamihara}}, \bibinfo {author} {\bibfnamefont {T.}~\bibnamefont {Watanabe}},
		\bibinfo {author} {\bibfnamefont {M.}~\bibnamefont {Hirano}},\ and\ \bibinfo
		{author} {\bibfnamefont {H.}~\bibnamefont {Hosono}},\ }\bibfield  {title}
	{\bibinfo {title} {Iron-based layered superconductor
			{{La[O$_{1-x}$F$_x$]FeAs}} (x=0.05-0.12) with {{$T_c = 26\,\mbox{K}$}}},\
	}\href {https://doi.org/10.1021/ja800073m} {\bibfield  {journal} {\bibinfo
			{journal} {Journal of the American Chemical Society}\ }\textbf {\bibinfo
			{volume} {130}},\ \bibinfo {pages} {3296} (\bibinfo {year}
		{2008})}\BibitemShut {NoStop}%
	\bibitem [{\citenamefont {Hsu}\ \emph {et~al.}(2008)\citenamefont {Hsu},
		\citenamefont {Luo}, \citenamefont {Yeh}, \citenamefont {Chen}, \citenamefont
		{Huang}, \citenamefont {Wu}, \citenamefont {Lee}, \citenamefont {Huang},
		\citenamefont {Chu}, \citenamefont {Yan},\ and\ \citenamefont
		{Wu}}]{SuperconductivityPbOtypeStructurehsu2008}%
	\BibitemOpen
	\bibfield  {author} {\bibinfo {author} {\bibfnamefont {F.-C.}\ \bibnamefont
			{Hsu}}, \bibinfo {author} {\bibfnamefont {J.-Y.}\ \bibnamefont {Luo}},
		\bibinfo {author} {\bibfnamefont {K.-W.}\ \bibnamefont {Yeh}}, \bibinfo
		{author} {\bibfnamefont {T.-K.}\ \bibnamefont {Chen}}, \bibinfo {author}
		{\bibfnamefont {T.-W.}\ \bibnamefont {Huang}}, \bibinfo {author}
		{\bibfnamefont {P.~M.}\ \bibnamefont {Wu}}, \bibinfo {author} {\bibfnamefont
			{Y.-C.}\ \bibnamefont {Lee}}, \bibinfo {author} {\bibfnamefont {Y.-L.}\
			\bibnamefont {Huang}}, \bibinfo {author} {\bibfnamefont {Y.-Y.}\ \bibnamefont
			{Chu}}, \bibinfo {author} {\bibfnamefont {D.-C.}\ \bibnamefont {Yan}},\ and\
		\bibinfo {author} {\bibfnamefont {M.-K.}\ \bibnamefont {Wu}},\ }\bibfield
	{title} {\bibinfo {title} {Superconductivity in the {{PbO-type}} structure
			$\alpha$-{{FeSe}}},\ }\href {https://doi.org/10.1073/pnas.0807325105}
	{\bibfield  {journal} {\bibinfo  {journal} {Proceedings of the National
				Academy of Sciences}\ }\textbf {\bibinfo {volume} {105}},\ \bibinfo {pages}
		{14262} (\bibinfo {year} {2008})}\BibitemShut {NoStop}%
	\bibitem [{\citenamefont {Margadonna}\ \emph {et~al.}(2009)\citenamefont
		{Margadonna}, \citenamefont {Takabayashi}, \citenamefont {Ohishi},
		\citenamefont {Mizuguchi}, \citenamefont {Takano}, \citenamefont {Kagayama},
		\citenamefont {Nakagawa}, \citenamefont {Takata},\ and\ \citenamefont
		{Prassides}}]{PhysRevB.80.064506}%
	\BibitemOpen
	\bibfield  {author} {\bibinfo {author} {\bibfnamefont {S.}~\bibnamefont
			{Margadonna}}, \bibinfo {author} {\bibfnamefont {Y.}~\bibnamefont
			{Takabayashi}}, \bibinfo {author} {\bibfnamefont {Y.}~\bibnamefont {Ohishi}},
		\bibinfo {author} {\bibfnamefont {Y.}~\bibnamefont {Mizuguchi}}, \bibinfo
		{author} {\bibfnamefont {Y.}~\bibnamefont {Takano}}, \bibinfo {author}
		{\bibfnamefont {T.}~\bibnamefont {Kagayama}}, \bibinfo {author}
		{\bibfnamefont {T.}~\bibnamefont {Nakagawa}}, \bibinfo {author}
		{\bibfnamefont {M.}~\bibnamefont {Takata}},\ and\ \bibinfo {author}
		{\bibfnamefont {K.}~\bibnamefont {Prassides}},\ }\bibfield  {title} {\bibinfo
		{title} {Pressure evolution of the low-temperature crystal structure and
			bonding of the superconductor {FeSe} $({T}_{c}=37\text{ }\text{K})$},\ }\href
	{https://doi.org/10.1103/PhysRevB.80.064506} {\bibfield  {journal} {\bibinfo
			{journal} {Phys. Rev. B}\ }\textbf {\bibinfo {volume} {80}},\ \bibinfo
		{pages} {064506} (\bibinfo {year} {2009})}\BibitemShut {NoStop}%
	\bibitem [{\citenamefont {Medvedev}\ \emph {et~al.}(2009)\citenamefont
		{Medvedev}, \citenamefont {McQueen}, \citenamefont {Troyan}, \citenamefont
		{Palasyuk}, \citenamefont {Eremets}, \citenamefont {Cava}, \citenamefont
		{Naghavi}, \citenamefont {Casper}, \citenamefont {Ksenofontov}, \citenamefont
		{Wortmann},\ and\ \citenamefont
		{Felser}}]{ElectronicMagneticPhasemedvedev2009}%
	\BibitemOpen
	\bibfield  {author} {\bibinfo {author} {\bibfnamefont {S.}~\bibnamefont
			{Medvedev}}, \bibinfo {author} {\bibfnamefont {T.~M.}\ \bibnamefont
			{McQueen}}, \bibinfo {author} {\bibfnamefont {I.~A.}\ \bibnamefont {Troyan}},
		\bibinfo {author} {\bibfnamefont {T.}~\bibnamefont {Palasyuk}}, \bibinfo
		{author} {\bibfnamefont {M.~I.}\ \bibnamefont {Eremets}}, \bibinfo {author}
		{\bibfnamefont {R.~J.}\ \bibnamefont {Cava}}, \bibinfo {author}
		{\bibfnamefont {S.}~\bibnamefont {Naghavi}}, \bibinfo {author} {\bibfnamefont
			{F.}~\bibnamefont {Casper}}, \bibinfo {author} {\bibfnamefont
			{V.}~\bibnamefont {Ksenofontov}}, \bibinfo {author} {\bibfnamefont
			{G.}~\bibnamefont {Wortmann}},\ and\ \bibinfo {author} {\bibfnamefont
			{C.}~\bibnamefont {Felser}},\ }\bibfield  {title} {\bibinfo {title}
		{Electronic and magnetic phase diagram of {$\beta$}-{{Fe$_{1.01}$}}{{Se}}
			with superconductivity at 36.7 {{K}} under pressure},\ }\href
	{https://doi.org/10.1038/nmat2491} {\bibfield  {journal} {\bibinfo  {journal}
			{Nature Materials}\ }\textbf {\bibinfo {volume} {8}},\ \bibinfo {pages} {630}
		(\bibinfo {year} {2009})}\BibitemShut {NoStop}%
	\bibitem [{\citenamefont {Kumar}\ \emph {et~al.}(2010)\citenamefont {Kumar},
		\citenamefont {Zhang}, \citenamefont {Sinogeikin}, \citenamefont {Xiao},
		\citenamefont {Kumar}, \citenamefont {Chow}, \citenamefont {Cornelius},\ and\
		\citenamefont {Chen}}]{CrystalElectronicStructurekumar2010a}%
	\BibitemOpen
	\bibfield  {author} {\bibinfo {author} {\bibfnamefont {R.~S.}\ \bibnamefont
			{Kumar}}, \bibinfo {author} {\bibfnamefont {Y.}~\bibnamefont {Zhang}},
		\bibinfo {author} {\bibfnamefont {S.}~\bibnamefont {Sinogeikin}}, \bibinfo
		{author} {\bibfnamefont {Y.}~\bibnamefont {Xiao}}, \bibinfo {author}
		{\bibfnamefont {S.}~\bibnamefont {Kumar}}, \bibinfo {author} {\bibfnamefont
			{P.}~\bibnamefont {Chow}}, \bibinfo {author} {\bibfnamefont {A.~L.}\
			\bibnamefont {Cornelius}},\ and\ \bibinfo {author} {\bibfnamefont
			{C.}~\bibnamefont {Chen}},\ }\bibfield  {title} {\bibinfo {title} {Crystal
			and {{Electronic Structure}} of {{FeSe}} at {{High Pressure}} and {{Low
					Temperature}}},\ }\href {https://doi.org/10.1021/jp1060446} {\bibfield
		{journal} {\bibinfo  {journal} {The Journal of Physical Chemistry B}\
		}\textbf {\bibinfo {volume} {114}},\ \bibinfo {pages} {12597} (\bibinfo
		{year} {2010})}\BibitemShut {NoStop}%
	\bibitem [{\citenamefont {Wang}\ \emph {et~al.}(2012)\citenamefont {Wang},
		\citenamefont {Li}, \citenamefont {Zhang}, \citenamefont {Zhang},
		\citenamefont {Zhang}, \citenamefont {Li}, \citenamefont {Ding},
		\citenamefont {Ou}, \citenamefont {Deng}, \citenamefont {Chang},
		\citenamefont {Wen}, \citenamefont {Song}, \citenamefont {He}, \citenamefont
		{Jia}, \citenamefont {Ji}, \citenamefont {Wang}, \citenamefont {Wang},
		\citenamefont {Chen}, \citenamefont {Ma},\ and\ \citenamefont
		{Xue}}]{InterfaceInducedHighTemperatureSuperconductivitywang2012}%
	\BibitemOpen
	\bibfield  {author} {\bibinfo {author} {\bibfnamefont {Q.-Y.}\ \bibnamefont
			{Wang}}, \bibinfo {author} {\bibfnamefont {Z.}~\bibnamefont {Li}}, \bibinfo
		{author} {\bibfnamefont {W.-H.}\ \bibnamefont {Zhang}}, \bibinfo {author}
		{\bibfnamefont {Z.-C.}\ \bibnamefont {Zhang}}, \bibinfo {author}
		{\bibfnamefont {J.-S.}\ \bibnamefont {Zhang}}, \bibinfo {author}
		{\bibfnamefont {W.}~\bibnamefont {Li}}, \bibinfo {author} {\bibfnamefont
			{H.}~\bibnamefont {Ding}}, \bibinfo {author} {\bibfnamefont {Y.-B.}\
			\bibnamefont {Ou}}, \bibinfo {author} {\bibfnamefont {P.}~\bibnamefont
			{Deng}}, \bibinfo {author} {\bibfnamefont {K.}~\bibnamefont {Chang}},
		\bibinfo {author} {\bibfnamefont {J.}~\bibnamefont {Wen}}, \bibinfo {author}
		{\bibfnamefont {C.-L.}\ \bibnamefont {Song}}, \bibinfo {author}
		{\bibfnamefont {K.}~\bibnamefont {He}}, \bibinfo {author} {\bibfnamefont
			{J.-F.}\ \bibnamefont {Jia}}, \bibinfo {author} {\bibfnamefont {S.-H.}\
			\bibnamefont {Ji}}, \bibinfo {author} {\bibfnamefont {Y.-Y.}\ \bibnamefont
			{Wang}}, \bibinfo {author} {\bibfnamefont {L.-L.}\ \bibnamefont {Wang}},
		\bibinfo {author} {\bibfnamefont {X.}~\bibnamefont {Chen}}, \bibinfo {author}
		{\bibfnamefont {X.-C.}\ \bibnamefont {Ma}},\ and\ \bibinfo {author}
		{\bibfnamefont {Q.-K.}\ \bibnamefont {Xue}},\ }\bibfield  {title} {\bibinfo
		{title} {Interface-{{Induced High-Temperature Superconductivity}} in {{Single
					Unit-Cell FeSe Films}} on {{SrTiO$_3$}}},\ }\href
	{https://doi.org/10.1088/0256-307X/29/3/037402} {\bibfield  {journal}
		{\bibinfo  {journal} {Chinese Physics Letters}\ }\textbf {\bibinfo {volume}
			{29}},\ \bibinfo {pages} {037402} (\bibinfo {year} {2012})}\BibitemShut
	{NoStop}%
	\bibitem [{\citenamefont {He}\ \emph {et~al.}(2013)\citenamefont {He},
		\citenamefont {He}, \citenamefont {Zhang}, \citenamefont {Zhao},
		\citenamefont {Liu}, \citenamefont {Liu}, \citenamefont {Mou}, \citenamefont
		{Ou}, \citenamefont {Wang}, \citenamefont {Li}, \citenamefont {Wang},
		\citenamefont {Peng}, \citenamefont {Liu}, \citenamefont {Chen},
		\citenamefont {Yu}, \citenamefont {Liu}, \citenamefont {Dong}, \citenamefont
		{Zhang}, \citenamefont {Chen}, \citenamefont {Xu}, \citenamefont {Chen},
		\citenamefont {Ma}, \citenamefont {Xue},\ and\ \citenamefont
		{Zhou}}]{he2013PhaseDiagramElectronica}%
	\BibitemOpen
	\bibfield  {author} {\bibinfo {author} {\bibfnamefont {S.}~\bibnamefont
			{He}}, \bibinfo {author} {\bibfnamefont {J.}~\bibnamefont {He}}, \bibinfo
		{author} {\bibfnamefont {W.}~\bibnamefont {Zhang}}, \bibinfo {author}
		{\bibfnamefont {L.}~\bibnamefont {Zhao}}, \bibinfo {author} {\bibfnamefont
			{D.}~\bibnamefont {Liu}}, \bibinfo {author} {\bibfnamefont {X.}~\bibnamefont
			{Liu}}, \bibinfo {author} {\bibfnamefont {D.}~\bibnamefont {Mou}}, \bibinfo
		{author} {\bibfnamefont {Y.-B.}\ \bibnamefont {Ou}}, \bibinfo {author}
		{\bibfnamefont {Q.-Y.}\ \bibnamefont {Wang}}, \bibinfo {author}
		{\bibfnamefont {Z.}~\bibnamefont {Li}}, \bibinfo {author} {\bibfnamefont
			{L.}~\bibnamefont {Wang}}, \bibinfo {author} {\bibfnamefont {Y.}~\bibnamefont
			{Peng}}, \bibinfo {author} {\bibfnamefont {Y.}~\bibnamefont {Liu}}, \bibinfo
		{author} {\bibfnamefont {C.}~\bibnamefont {Chen}}, \bibinfo {author}
		{\bibfnamefont {L.}~\bibnamefont {Yu}}, \bibinfo {author} {\bibfnamefont
			{G.}~\bibnamefont {Liu}}, \bibinfo {author} {\bibfnamefont {X.}~\bibnamefont
			{Dong}}, \bibinfo {author} {\bibfnamefont {J.}~\bibnamefont {Zhang}},
		\bibinfo {author} {\bibfnamefont {C.}~\bibnamefont {Chen}}, \bibinfo {author}
		{\bibfnamefont {Z.}~\bibnamefont {Xu}}, \bibinfo {author} {\bibfnamefont
			{X.}~\bibnamefont {Chen}}, \bibinfo {author} {\bibfnamefont {X.}~\bibnamefont
			{Ma}}, \bibinfo {author} {\bibfnamefont {Q.}~\bibnamefont {Xue}},\ and\
		\bibinfo {author} {\bibfnamefont {X.~J.}\ \bibnamefont {Zhou}},\ }\bibfield
	{title} {\bibinfo {title} {Phase diagram and electronic indication of
			high-temperature superconductivity at 65 {{K}} in single-layer {{FeSe}}
			films},\ }\href {https://doi.org/10.1038/nmat3648} {\bibfield  {journal}
		{\bibinfo  {journal} {Nature Materials}\ }\textbf {\bibinfo {volume} {12}},\
		\bibinfo {pages} {605} (\bibinfo {year} {2013})}\BibitemShut {NoStop}%
	\bibitem [{\citenamefont {Ge}\ \emph {et~al.}(2015)\citenamefont {Ge},
		\citenamefont {Liu}, \citenamefont {Liu}, \citenamefont {Gao}, \citenamefont
		{Qian}, \citenamefont {Xue}, \citenamefont {Liu},\ and\ \citenamefont
		{Jia}}]{Ge2015}%
	\BibitemOpen
	\bibfield  {author} {\bibinfo {author} {\bibfnamefont {J.-F.}\ \bibnamefont
			{Ge}}, \bibinfo {author} {\bibfnamefont {Z.-L.}\ \bibnamefont {Liu}},
		\bibinfo {author} {\bibfnamefont {C.}~\bibnamefont {Liu}}, \bibinfo {author}
		{\bibfnamefont {C.-L.}\ \bibnamefont {Gao}}, \bibinfo {author} {\bibfnamefont
			{D.}~\bibnamefont {Qian}}, \bibinfo {author} {\bibfnamefont {Q.-K.}\
			\bibnamefont {Xue}}, \bibinfo {author} {\bibfnamefont {Y.}~\bibnamefont
			{Liu}},\ and\ \bibinfo {author} {\bibfnamefont {J.-F.}\ \bibnamefont {Jia}},\
	}\bibfield  {title} {\bibinfo {title} {Superconductivity above {100 K} in
			single-layer {FeSe} films on doped {SrTiO$_3$}},\ }\href
	{https://doi.org/10.1038/nmat4153} {\bibfield  {journal} {\bibinfo  {journal}
			{Nature Materials}\ }\textbf {\bibinfo {volume} {14}},\ \bibinfo {pages}
		{285} (\bibinfo {year} {2015})}\BibitemShut {NoStop}%
	\bibitem [{\citenamefont {Guo}\ \emph {et~al.}(2010)\citenamefont {Guo},
		\citenamefont {Jin}, \citenamefont {Wang}, \citenamefont {Wang},
		\citenamefont {Zhu}, \citenamefont {Zhou}, \citenamefont {He},\ and\
		\citenamefont {Chen}}]{SuperconductivityIronSelenideguo2010}%
	\BibitemOpen
	\bibfield  {author} {\bibinfo {author} {\bibfnamefont {J.}~\bibnamefont
			{Guo}}, \bibinfo {author} {\bibfnamefont {S.}~\bibnamefont {Jin}}, \bibinfo
		{author} {\bibfnamefont {G.}~\bibnamefont {Wang}}, \bibinfo {author}
		{\bibfnamefont {S.}~\bibnamefont {Wang}}, \bibinfo {author} {\bibfnamefont
			{K.}~\bibnamefont {Zhu}}, \bibinfo {author} {\bibfnamefont {T.}~\bibnamefont
			{Zhou}}, \bibinfo {author} {\bibfnamefont {M.}~\bibnamefont {He}},\ and\
		\bibinfo {author} {\bibfnamefont {X.}~\bibnamefont {Chen}},\ }\bibfield
	{title} {\bibinfo {title} {Superconductivity in the iron selenide
			{{K$_x$Fe$_2$Se$_2$}} {{($0\le x \le 1.0$)}}},\ }\href
	{https://doi.org/10.1103/PhysRevB.82.180520} {\bibfield  {journal} {\bibinfo
			{journal} {Phys. Rev. B}\ }\textbf {\bibinfo {volume} {82}},\ \bibinfo
		{pages} {180520(R)} (\bibinfo {year} {2010})}\BibitemShut {NoStop}%
	\bibitem [{\citenamefont {Liu}\ \emph {et~al.}(2011)\citenamefont {Liu},
		\citenamefont {Luo}, \citenamefont {Zhang}, \citenamefont {Wang},
		\citenamefont {Ying}, \citenamefont {Wang}, \citenamefont {Yan},
		\citenamefont {Xiang}, \citenamefont {Cheng}, \citenamefont {Ye},
		\citenamefont {Li},\ and\ \citenamefont
		{Chen}}]{CoexistenceSuperconductivityAntiferromagnetismliu2011}%
	\BibitemOpen
	\bibfield  {author} {\bibinfo {author} {\bibfnamefont {R.~H.}\ \bibnamefont
			{Liu}}, \bibinfo {author} {\bibfnamefont {X.~G.}\ \bibnamefont {Luo}},
		\bibinfo {author} {\bibfnamefont {M.}~\bibnamefont {Zhang}}, \bibinfo
		{author} {\bibfnamefont {A.~F.}\ \bibnamefont {Wang}}, \bibinfo {author}
		{\bibfnamefont {J.~J.}\ \bibnamefont {Ying}}, \bibinfo {author}
		{\bibfnamefont {X.~F.}\ \bibnamefont {Wang}}, \bibinfo {author}
		{\bibfnamefont {Y.~J.}\ \bibnamefont {Yan}}, \bibinfo {author} {\bibfnamefont
			{Z.~J.}\ \bibnamefont {Xiang}}, \bibinfo {author} {\bibfnamefont
			{P.}~\bibnamefont {Cheng}}, \bibinfo {author} {\bibfnamefont {G.~J.}\
			\bibnamefont {Ye}}, \bibinfo {author} {\bibfnamefont {Z.~Y.}\ \bibnamefont
			{Li}},\ and\ \bibinfo {author} {\bibfnamefont {X.~H.}\ \bibnamefont {Chen}},\
	}\bibfield  {title} {\bibinfo {title} {Coexistence of superconductivity and
			antiferromagnetism in single crystals
			{{A}}{\textsubscript{0.8}}{{Fe}}{\textsubscript{2-y}}{{Se}}{\textsubscript{2}}
			({{A}}={{K}}, {{Rb}}, {{Cs}}, {{Tl}}/{{K}} and {{Tl}}/{{Rb}}): {{Evidence}}
			from magnetization and resistivity},\ }\href
	{https://doi.org/10.1209/0295-5075/94/27008} {\bibfield  {journal} {\bibinfo
			{journal} {EPL (Europhysics Letters)}\ }\textbf {\bibinfo {volume} {94}},\
		\bibinfo {pages} {27008} (\bibinfo {year} {2011})}\BibitemShut {NoStop}%
	\bibitem [{\citenamefont {Chen}\ \emph {et~al.}(2011)\citenamefont {Chen},
		\citenamefont {Xu}, \citenamefont {Ge}, \citenamefont {Zhang}, \citenamefont
		{Ye}, \citenamefont {Yang}, \citenamefont {Jiang}, \citenamefont {Xie},
		\citenamefont {Che}, \citenamefont {Zhang}, \citenamefont {Wang},
		\citenamefont {Chen}, \citenamefont {Shen}, \citenamefont {Hu},\ and\
		\citenamefont {Feng}}]{Chen2011}%
	\BibitemOpen
	\bibfield  {author} {\bibinfo {author} {\bibfnamefont {F.}~\bibnamefont
			{Chen}}, \bibinfo {author} {\bibfnamefont {M.}~\bibnamefont {Xu}}, \bibinfo
		{author} {\bibfnamefont {Q.~Q.}\ \bibnamefont {Ge}}, \bibinfo {author}
		{\bibfnamefont {Y.}~\bibnamefont {Zhang}}, \bibinfo {author} {\bibfnamefont
			{Z.~R.}\ \bibnamefont {Ye}}, \bibinfo {author} {\bibfnamefont {L.~X.}\
			\bibnamefont {Yang}}, \bibinfo {author} {\bibfnamefont {J.}~\bibnamefont
			{Jiang}}, \bibinfo {author} {\bibfnamefont {B.~P.}\ \bibnamefont {Xie}},
		\bibinfo {author} {\bibfnamefont {R.~C.}\ \bibnamefont {Che}}, \bibinfo
		{author} {\bibfnamefont {M.}~\bibnamefont {Zhang}}, \bibinfo {author}
		{\bibfnamefont {A.~F.}\ \bibnamefont {Wang}}, \bibinfo {author}
		{\bibfnamefont {X.~H.}\ \bibnamefont {Chen}}, \bibinfo {author}
		{\bibfnamefont {D.~W.}\ \bibnamefont {Shen}}, \bibinfo {author}
		{\bibfnamefont {J.~P.}\ \bibnamefont {Hu}},\ and\ \bibinfo {author}
		{\bibfnamefont {D.~L.}\ \bibnamefont {Feng}},\ }\bibfield  {title} {\bibinfo
		{title} {Electronic identification of the parental phases and mesoscopic
			phase separation of
			{${\mathrm{K}}_{x}{\mathrm{Fe}}_{2\ensuremath{-}y}{\mathrm{Se}}_{2}$}
			superconductors},\ }\href {https://doi.org/10.1103/PhysRevX.1.021020}
	{\bibfield  {journal} {\bibinfo  {journal} {Phys. Rev. X}\ }\textbf {\bibinfo
			{volume} {1}},\ \bibinfo {pages} {021020} (\bibinfo {year}
		{2011})}\BibitemShut {NoStop}%
	\bibitem [{\citenamefont {Texier}\ \emph {et~al.}(2012)\citenamefont {Texier},
		\citenamefont {Deisenhofer}, \citenamefont {Tsurkan}, \citenamefont {Loidl},
		\citenamefont {Inosov}, \citenamefont {Friemel},\ and\ \citenamefont
		{Bobroff}}]{Texier2012}%
	\BibitemOpen
	\bibfield  {author} {\bibinfo {author} {\bibfnamefont {Y.}~\bibnamefont
			{Texier}}, \bibinfo {author} {\bibfnamefont {J.}~\bibnamefont {Deisenhofer}},
		\bibinfo {author} {\bibfnamefont {V.}~\bibnamefont {Tsurkan}}, \bibinfo
		{author} {\bibfnamefont {A.}~\bibnamefont {Loidl}}, \bibinfo {author}
		{\bibfnamefont {D.~S.}\ \bibnamefont {Inosov}}, \bibinfo {author}
		{\bibfnamefont {G.}~\bibnamefont {Friemel}},\ and\ \bibinfo {author}
		{\bibfnamefont {J.}~\bibnamefont {Bobroff}},\ }\bibfield  {title} {\bibinfo
		{title} {{NMR} study in the iron-selenide
			{${\mathrm{Rb}}_{0.74}{\mathrm{Fe}}_{1.6}{\mathrm{Se}}_{2}$: D}etermination
			of the superconducting phase as iron vacancy-free
			{${\mathrm{Rb}}_{0.3}{\mathrm{Fe}}_{2}{\mathrm{Se}}_{2}$}},\ }\href
	{https://doi.org/10.1103/PhysRevLett.108.237002} {\bibfield  {journal}
		{\bibinfo  {journal} {Phys. Rev. Lett.}\ }\textbf {\bibinfo {volume} {108}},\
		\bibinfo {pages} {237002} (\bibinfo {year} {2012})}\BibitemShut {NoStop}%
	\bibitem [{\citenamefont {Li}\ \emph {et~al.}(2012)\citenamefont {Li},
		\citenamefont {Ding}, \citenamefont {Deng}, \citenamefont {Chang},
		\citenamefont {Song}, \citenamefont {He}, \citenamefont {Wang}, \citenamefont
		{Ma}, \citenamefont {Hu}, \citenamefont {Chen},\ and\ \citenamefont
		{Xue}}]{Li2021}%
	\BibitemOpen
	\bibfield  {author} {\bibinfo {author} {\bibfnamefont {W.}~\bibnamefont
			{Li}}, \bibinfo {author} {\bibfnamefont {H.}~\bibnamefont {Ding}}, \bibinfo
		{author} {\bibfnamefont {P.}~\bibnamefont {Deng}}, \bibinfo {author}
		{\bibfnamefont {K.}~\bibnamefont {Chang}}, \bibinfo {author} {\bibfnamefont
			{C.}~\bibnamefont {Song}}, \bibinfo {author} {\bibfnamefont {K.}~\bibnamefont
			{He}}, \bibinfo {author} {\bibfnamefont {L.}~\bibnamefont {Wang}}, \bibinfo
		{author} {\bibfnamefont {X.}~\bibnamefont {Ma}}, \bibinfo {author}
		{\bibfnamefont {J.-P.}\ \bibnamefont {Hu}}, \bibinfo {author} {\bibfnamefont
			{X.}~\bibnamefont {Chen}},\ and\ \bibinfo {author} {\bibfnamefont {Q.-K.}\
			\bibnamefont {Xue}},\ }\bibfield  {title} {\bibinfo {title} {Phase separation
			and magnetic order in {K}-doped iron selenide superconductor},\ }\href
	{https://doi.org/10.1038/nphys2155} {\bibfield  {journal} {\bibinfo
			{journal} {Nature Physics}\ }\textbf {\bibinfo {volume} {8}},\ \bibinfo
		{pages} {126} (\bibinfo {year} {2012})}\BibitemShut {NoStop}%
	\bibitem [{\citenamefont {Sedlmaier}\ \emph {et~al.}(2014)\citenamefont
		{Sedlmaier}, \citenamefont {Cassidy}, \citenamefont {Morris}, \citenamefont
		{Drakopoulos}, \citenamefont {Reinhard}, \citenamefont {Moorhouse},
		\citenamefont {O’Hare}, \citenamefont {Manuel}, \citenamefont {Khalyavin},\
		and\ \citenamefont
		{Clarke}}]{AmmoniaRichHighTemperatureSuperconductingsedlmaier2014}%
	\BibitemOpen
	\bibfield  {author} {\bibinfo {author} {\bibfnamefont {S.~J.}\ \bibnamefont
			{Sedlmaier}}, \bibinfo {author} {\bibfnamefont {S.~J.}\ \bibnamefont
			{Cassidy}}, \bibinfo {author} {\bibfnamefont {R.~G.}\ \bibnamefont {Morris}},
		\bibinfo {author} {\bibfnamefont {M.}~\bibnamefont {Drakopoulos}}, \bibinfo
		{author} {\bibfnamefont {C.}~\bibnamefont {Reinhard}}, \bibinfo {author}
		{\bibfnamefont {S.~J.}\ \bibnamefont {Moorhouse}}, \bibinfo {author}
		{\bibfnamefont {D.}~\bibnamefont {O’Hare}}, \bibinfo {author}
		{\bibfnamefont {P.}~\bibnamefont {Manuel}}, \bibinfo {author} {\bibfnamefont
			{D.}~\bibnamefont {Khalyavin}},\ and\ \bibinfo {author} {\bibfnamefont
			{S.~J.}\ \bibnamefont {Clarke}},\ }\bibfield  {title} {\bibinfo {title}
		{Ammonia-{{Rich High-Temperature Superconducting Intercalates}} of {{Iron
					Selenide Revealed}} through {{Time-Resolved}} {\emph{in }}{{{\emph{Situ}}}}
			{{X-ray}} and {{Neutron Diffraction}}},\ }\href
	{https://doi.org/10.1021/ja411624q} {\bibfield  {journal} {\bibinfo
			{journal} {Journal of the American Chemical Society}\ }\textbf {\bibinfo
			{volume} {136}},\ \bibinfo {pages} {630} (\bibinfo {year}
		{2014})}\BibitemShut {NoStop}%
	\bibitem [{\citenamefont {Burrard-Lucas}\ \emph {et~al.}(2013)\citenamefont
		{Burrard-Lucas}, \citenamefont {Free}, \citenamefont {Sedlmaier},
		\citenamefont {Wright}, \citenamefont {Cassidy}, \citenamefont {Hara},
		\citenamefont {Corkett}, \citenamefont {Lancaster}, \citenamefont {Baker},
		\citenamefont {Blundell},\ and\ \citenamefont
		{Clarke}}]{EnhancementSuperconductingTransitionburrard-lucas2013}%
	\BibitemOpen
	\bibfield  {author} {\bibinfo {author} {\bibfnamefont {M.}~\bibnamefont
			{Burrard-Lucas}}, \bibinfo {author} {\bibfnamefont {D.~G.}\ \bibnamefont
			{Free}}, \bibinfo {author} {\bibfnamefont {S.~J.}\ \bibnamefont {Sedlmaier}},
		\bibinfo {author} {\bibfnamefont {J.~D.}\ \bibnamefont {Wright}}, \bibinfo
		{author} {\bibfnamefont {S.~J.}\ \bibnamefont {Cassidy}}, \bibinfo {author}
		{\bibfnamefont {Y.}~\bibnamefont {Hara}}, \bibinfo {author} {\bibfnamefont
			{A.~J.}\ \bibnamefont {Corkett}}, \bibinfo {author} {\bibfnamefont
			{T.}~\bibnamefont {Lancaster}}, \bibinfo {author} {\bibfnamefont {P.~J.}\
			\bibnamefont {Baker}}, \bibinfo {author} {\bibfnamefont {S.~J.}\ \bibnamefont
			{Blundell}},\ and\ \bibinfo {author} {\bibfnamefont {S.~J.}\ \bibnamefont
			{Clarke}},\ }\bibfield  {title} {\bibinfo {title} {Enhancement of the
			superconducting transition temperature of {{FeSe}} by intercalation of a
			molecular spacer layer},\ }\href {https://doi.org/10.1038/nmat3464}
	{\bibfield  {journal} {\bibinfo  {journal} {Nature Materials}\ }\textbf
		{\bibinfo {volume} {12}},\ \bibinfo {pages} {15} (\bibinfo {year}
		{2013})}\BibitemShut {NoStop}%
	\bibitem [{\citenamefont {Sun}\ \emph {et~al.}(2015)\citenamefont {Sun},
		\citenamefont {Woodruff}, \citenamefont {Cassidy}, \citenamefont {Allcroft},
		\citenamefont {Sedlmaier}, \citenamefont {Thompson}, \citenamefont {Bingham},
		\citenamefont {Forder}, \citenamefont {Cartenet}, \citenamefont {Mary},
		\citenamefont {Ramos}, \citenamefont {Foronda}, \citenamefont {Williams},
		\citenamefont {Li}, \citenamefont {Blundell},\ and\ \citenamefont
		{Clarke}}]{SoftChemicalControlsun2015}%
	\BibitemOpen
	\bibfield  {author} {\bibinfo {author} {\bibfnamefont {H.}~\bibnamefont
			{Sun}}, \bibinfo {author} {\bibfnamefont {D.~N.}\ \bibnamefont {Woodruff}},
		\bibinfo {author} {\bibfnamefont {S.~J.}\ \bibnamefont {Cassidy}}, \bibinfo
		{author} {\bibfnamefont {G.~M.}\ \bibnamefont {Allcroft}}, \bibinfo {author}
		{\bibfnamefont {S.~J.}\ \bibnamefont {Sedlmaier}}, \bibinfo {author}
		{\bibfnamefont {A.~L.}\ \bibnamefont {Thompson}}, \bibinfo {author}
		{\bibfnamefont {P.~A.}\ \bibnamefont {Bingham}}, \bibinfo {author}
		{\bibfnamefont {S.~D.}\ \bibnamefont {Forder}}, \bibinfo {author}
		{\bibfnamefont {S.}~\bibnamefont {Cartenet}}, \bibinfo {author}
		{\bibfnamefont {N.}~\bibnamefont {Mary}}, \bibinfo {author} {\bibfnamefont
			{S.}~\bibnamefont {Ramos}}, \bibinfo {author} {\bibfnamefont {F.~R.}\
			\bibnamefont {Foronda}}, \bibinfo {author} {\bibfnamefont {B.~H.}\
			\bibnamefont {Williams}}, \bibinfo {author} {\bibfnamefont {X.}~\bibnamefont
			{Li}}, \bibinfo {author} {\bibfnamefont {S.~J.}\ \bibnamefont {Blundell}},\
		and\ \bibinfo {author} {\bibfnamefont {S.~J.}\ \bibnamefont {Clarke}},\
	}\bibfield  {title} {\bibinfo {title} {Soft {{Chemical Control}} of
			{{Superconductivity}} in {{Lithium Iron Selenide Hydroxides
					Li}}{\textsubscript{1–{\emph{x}}}}{{Fe}}{\textsubscript{{\emph{x}}}}({{OH}}){{Fe}}{\textsubscript{1–{\emph{y}}}}{{Se}}},\
	}\href {https://doi.org/10.1021/ic5028702} {\bibfield  {journal} {\bibinfo
			{journal} {Inorganic Chemistry}\ }\textbf {\bibinfo {volume} {54}},\ \bibinfo
		{pages} {1958} (\bibinfo {year} {2015})}\BibitemShut {NoStop}%
	\bibitem [{\citenamefont {Lu}\ \emph {et~al.}(2015)\citenamefont {Lu},
		\citenamefont {Wang}, \citenamefont {Wu}, \citenamefont {Wu}, \citenamefont
		{Zhao}, \citenamefont {Zeng}, \citenamefont {Luo}, \citenamefont {Wu},
		\citenamefont {Bao}, \citenamefont {Zhang}, \citenamefont {Huang},
		\citenamefont {Huang},\ and\ \citenamefont {Chen}}]{Lu2015}%
	\BibitemOpen
	\bibfield  {author} {\bibinfo {author} {\bibfnamefont {X.~F.}\ \bibnamefont
			{Lu}}, \bibinfo {author} {\bibfnamefont {N.~Z.}\ \bibnamefont {Wang}},
		\bibinfo {author} {\bibfnamefont {H.}~\bibnamefont {Wu}}, \bibinfo {author}
		{\bibfnamefont {Y.~P.}\ \bibnamefont {Wu}}, \bibinfo {author} {\bibfnamefont
			{D.}~\bibnamefont {Zhao}}, \bibinfo {author} {\bibfnamefont {X.~Z.}\
			\bibnamefont {Zeng}}, \bibinfo {author} {\bibfnamefont {X.~G.}\ \bibnamefont
			{Luo}}, \bibinfo {author} {\bibfnamefont {T.}~\bibnamefont {Wu}}, \bibinfo
		{author} {\bibfnamefont {W.}~\bibnamefont {Bao}}, \bibinfo {author}
		{\bibfnamefont {G.~H.}\ \bibnamefont {Zhang}}, \bibinfo {author}
		{\bibfnamefont {F.~Q.}\ \bibnamefont {Huang}}, \bibinfo {author}
		{\bibfnamefont {Q.~Z.}\ \bibnamefont {Huang}},\ and\ \bibinfo {author}
		{\bibfnamefont {X.~H.}\ \bibnamefont {Chen}},\ }\bibfield  {title} {\bibinfo
		{title} {Coexistence of superconductivity and antiferromagnetism in
			{(Li$_{0.8}$Fe$_{0.8}$)OHFeSe}},\ }\href {https://doi.org/10.1038/nmat4155}
	{\bibfield  {journal} {\bibinfo  {journal} {Nature materials}\ }\textbf
		{\bibinfo {volume} {14}},\ \bibinfo {pages} {325} (\bibinfo {year}
		{2015})}\BibitemShut {NoStop}%
	\bibitem [{\citenamefont {Kuroki}\ \emph {et~al.}(2008)\citenamefont {Kuroki},
		\citenamefont {Onari}, \citenamefont {Arita}, \citenamefont {Usui},
		\citenamefont {Tanaka}, \citenamefont {Kontani},\ and\ \citenamefont
		{Aoki}}]{UnconventionalPairingOriginatingkuroki2008}%
	\BibitemOpen
	\bibfield  {author} {\bibinfo {author} {\bibfnamefont {K.}~\bibnamefont
			{Kuroki}}, \bibinfo {author} {\bibfnamefont {S.}~\bibnamefont {Onari}},
		\bibinfo {author} {\bibfnamefont {R.}~\bibnamefont {Arita}}, \bibinfo
		{author} {\bibfnamefont {H.}~\bibnamefont {Usui}}, \bibinfo {author}
		{\bibfnamefont {Y.}~\bibnamefont {Tanaka}}, \bibinfo {author} {\bibfnamefont
			{H.}~\bibnamefont {Kontani}},\ and\ \bibinfo {author} {\bibfnamefont
			{H.}~\bibnamefont {Aoki}},\ }\bibfield  {title} {\bibinfo {title}
		{Unconventional {{Pairing Originating}} from the {{Disconnected Fermi
					Surfaces}} of {{Superconducting LaFeAsO}}$_{1-x}${{F}}$_x$},\ }\href
	{https://doi.org/10.1103/PhysRevLett.101.087004} {\bibfield  {journal}
		{\bibinfo  {journal} {Phys. Rev. Lett.}\ }\textbf {\bibinfo {volume} {101}},\
		\bibinfo {pages} {087004} (\bibinfo {year} {2008})}\BibitemShut {NoStop}%
	\bibitem [{\citenamefont {Mazin}\ \emph {et~al.}(2008)\citenamefont {Mazin},
		\citenamefont {Singh}, \citenamefont {Johannes},\ and\ \citenamefont
		{Du}}]{UnconventionalSuperconductivitywithaSignReversalintheOrder}%
	\BibitemOpen
	\bibfield  {author} {\bibinfo {author} {\bibfnamefont {I.~I.}\ \bibnamefont
			{Mazin}}, \bibinfo {author} {\bibfnamefont {D.~J.}\ \bibnamefont {Singh}},
		\bibinfo {author} {\bibfnamefont {M.~D.}\ \bibnamefont {Johannes}},\ and\
		\bibinfo {author} {\bibfnamefont {M.~H.}\ \bibnamefont {Du}},\ }\bibfield
	{title} {\bibinfo {title} {Unconventional superconductivity with a sign
			reversal in the order parameter of {{LaFeAsO$_{1-x}$F$_{x}$}}},\ }\href
	{https://doi.org/10.1103/PhysRevLett.101.057003} {\bibfield  {journal}
		{\bibinfo  {journal} {Phys. Rev. Lett.}\ }\textbf {\bibinfo {volume} {101}},\
		\bibinfo {pages} {057003} (\bibinfo {year} {2008})}\BibitemShut {NoStop}%
	\bibitem [{\citenamefont {Mazin}(2011)}]{Mazin2011}%
	\BibitemOpen
	\bibfield  {author} {\bibinfo {author} {\bibfnamefont {I.~I.}\ \bibnamefont
			{Mazin}},\ }\bibfield  {title} {\bibinfo {title} {Symmetry analysis of
			possible superconducting states in {K${}_{x}$Fe${}_{y}$Se${}_{2}$}
			superconductors},\ }\href {https://doi.org/10.1103/PhysRevB.84.024529}
	{\bibfield  {journal} {\bibinfo  {journal} {Phys. Rev. B}\ }\textbf {\bibinfo
			{volume} {84}},\ \bibinfo {pages} {024529} (\bibinfo {year}
		{2011})}\BibitemShut {NoStop}%
	\bibitem [{\citenamefont {Wang}\ \emph {et~al.}(2017)\citenamefont {Wang},
		\citenamefont {Hardy}, \citenamefont {Wolf}, \citenamefont {Adelmann},
		\citenamefont {Fromknecht}, \citenamefont {Schweiss},\ and\ \citenamefont
		{Meingast}}]{SuperconductivityenhancedNematicityGapwang2017}%
	\BibitemOpen
	\bibfield  {author} {\bibinfo {author} {\bibfnamefont {L.}~\bibnamefont
			{Wang}}, \bibinfo {author} {\bibfnamefont {F.}~\bibnamefont {Hardy}},
		\bibinfo {author} {\bibfnamefont {T.}~\bibnamefont {Wolf}}, \bibinfo {author}
		{\bibfnamefont {P.}~\bibnamefont {Adelmann}}, \bibinfo {author}
		{\bibfnamefont {R.}~\bibnamefont {Fromknecht}}, \bibinfo {author}
		{\bibfnamefont {P.}~\bibnamefont {Schweiss}},\ and\ \bibinfo {author}
		{\bibfnamefont {C.}~\bibnamefont {Meingast}},\ }\bibfield  {title} {\bibinfo
		{title} {Superconductivity-enhanced nematicity and “s+d” gap symmetry in
			{{Fe}}({{Se$_{1-x}$}}{{S$_x$}})},\ }\href
	{https://doi.org/10.1002/pssb.201600153} {\bibfield  {journal} {\bibinfo
			{journal} {physica status solidi (b)}\ }\textbf {\bibinfo {volume} {254}},\
		\bibinfo {pages} {1600153} (\bibinfo {year} {2017})}\BibitemShut {NoStop}%
	\bibitem [{\citenamefont {Kasahara}\ \emph {et~al.}(2014)\citenamefont
		{Kasahara}, \citenamefont {Watashige}, \citenamefont {Hanaguri},
		\citenamefont {Kohsaka}, \citenamefont {Yamashita}, \citenamefont
		{Shimoyama}, \citenamefont {Mizukami}, \citenamefont {Endo}, \citenamefont
		{Ikeda}, \citenamefont {Aoyama} \emph
		{et~al.}}]{FieldinducedSuperconductingPhasekasahara2014}%
	\BibitemOpen
	\bibfield  {author} {\bibinfo {author} {\bibfnamefont {S.}~\bibnamefont
			{Kasahara}}, \bibinfo {author} {\bibfnamefont {T.}~\bibnamefont {Watashige}},
		\bibinfo {author} {\bibfnamefont {T.}~\bibnamefont {Hanaguri}}, \bibinfo
		{author} {\bibfnamefont {Y.}~\bibnamefont {Kohsaka}}, \bibinfo {author}
		{\bibfnamefont {T.}~\bibnamefont {Yamashita}}, \bibinfo {author}
		{\bibfnamefont {Y.}~\bibnamefont {Shimoyama}}, \bibinfo {author}
		{\bibfnamefont {Y.}~\bibnamefont {Mizukami}}, \bibinfo {author}
		{\bibfnamefont {R.}~\bibnamefont {Endo}}, \bibinfo {author} {\bibfnamefont
			{H.}~\bibnamefont {Ikeda}}, \bibinfo {author} {\bibfnamefont
			{K.}~\bibnamefont {Aoyama}}, \emph {et~al.},\ }\bibfield  {title} {\bibinfo
		{title} {Field-induced superconducting phase of {{FeSe}} in the {{BCS-BEC}}
			cross-over},\ }\href {https://doi.org/10.1073/pnas.1413477111} {\bibfield
		{journal} {\bibinfo  {journal} {Proceedings of the National Academy of
				Sciences}\ }\textbf {\bibinfo {volume} {111}},\ \bibinfo {pages} {16309}
		(\bibinfo {year} {2014})}\BibitemShut {NoStop}%
	\bibitem [{\citenamefont {Biswas}\ \emph {et~al.}(2018)\citenamefont {Biswas},
		\citenamefont {Kreisel}, \citenamefont {Wang}, \citenamefont {Adroja},
		\citenamefont {Hillier}, \citenamefont {Zhao}, \citenamefont {Khasanov},
		\citenamefont {Orain}, \citenamefont {Amato},\ and\ \citenamefont
		{Morenzoni}}]{biswas2018EvidenceNodalGap}%
	\BibitemOpen
	\bibfield  {author} {\bibinfo {author} {\bibfnamefont {P.~K.}\ \bibnamefont
			{Biswas}}, \bibinfo {author} {\bibfnamefont {A.}~\bibnamefont {Kreisel}},
		\bibinfo {author} {\bibfnamefont {Q.}~\bibnamefont {Wang}}, \bibinfo {author}
		{\bibfnamefont {D.~T.}\ \bibnamefont {Adroja}}, \bibinfo {author}
		{\bibfnamefont {A.~D.}\ \bibnamefont {Hillier}}, \bibinfo {author}
		{\bibfnamefont {J.}~\bibnamefont {Zhao}}, \bibinfo {author} {\bibfnamefont
			{R.}~\bibnamefont {Khasanov}}, \bibinfo {author} {\bibfnamefont {J.-C.}\
			\bibnamefont {Orain}}, \bibinfo {author} {\bibfnamefont {A.}~\bibnamefont
			{Amato}},\ and\ \bibinfo {author} {\bibfnamefont {E.}~\bibnamefont
			{Morenzoni}},\ }\bibfield  {title} {\bibinfo {title} {Evidence of nodal gap
			structure in the basal plane of the {{FeSe}} superconductor},\ }\href
	{https://doi.org/10.1103/PhysRevB.98.180501} {\bibfield  {journal} {\bibinfo
			{journal} {Phys. Rev. B}\ }\textbf {\bibinfo {volume} {98}},\ \bibinfo
		{pages} {180501(R)} (\bibinfo {year} {2018})}\BibitemShut {NoStop}%
	\bibitem [{\citenamefont {Khasanov}\ \emph {et~al.}(2008)\citenamefont
		{Khasanov}, \citenamefont {Conder}, \citenamefont {Pomjakushina},
		\citenamefont {Amato}, \citenamefont {Baines}, \citenamefont {Bukowski},
		\citenamefont {Karpinski}, \citenamefont {Katrych}, \citenamefont {Klauss},
		\citenamefont {Luetkens}, \citenamefont {Shengelaya},\ and\ \citenamefont
		{Zhigadlo}}]{khasanov2008EvidenceNodelessSuperconductivity}%
	\BibitemOpen
	\bibfield  {author} {\bibinfo {author} {\bibfnamefont {R.}~\bibnamefont
			{Khasanov}}, \bibinfo {author} {\bibfnamefont {K.}~\bibnamefont {Conder}},
		\bibinfo {author} {\bibfnamefont {E.}~\bibnamefont {Pomjakushina}}, \bibinfo
		{author} {\bibfnamefont {A.}~\bibnamefont {Amato}}, \bibinfo {author}
		{\bibfnamefont {C.}~\bibnamefont {Baines}}, \bibinfo {author} {\bibfnamefont
			{Z.}~\bibnamefont {Bukowski}}, \bibinfo {author} {\bibfnamefont
			{J.}~\bibnamefont {Karpinski}}, \bibinfo {author} {\bibfnamefont
			{S.}~\bibnamefont {Katrych}}, \bibinfo {author} {\bibfnamefont {H.-H.}\
			\bibnamefont {Klauss}}, \bibinfo {author} {\bibfnamefont {H.}~\bibnamefont
			{Luetkens}}, \bibinfo {author} {\bibfnamefont {A.}~\bibnamefont
			{Shengelaya}},\ and\ \bibinfo {author} {\bibfnamefont {N.~D.}\ \bibnamefont
			{Zhigadlo}},\ }\bibfield  {title} {\bibinfo {title} {Evidence of nodeless
			superconductivity in {{FeSe$_{0.85}$}} from a muon-spin-rotation study of the
			in-plane magnetic penetration depth},\ }\href
	{https://doi.org/10.1103/PhysRevB.78.220510} {\bibfield  {journal} {\bibinfo
			{journal} {Phys. Rev. B}\ }\textbf {\bibinfo {volume} {78}},\ \bibinfo
		{pages} {220510(R)} (\bibinfo {year} {2008})}\BibitemShut {NoStop}%
	\bibitem [{\citenamefont {Chen}\ \emph {et~al.}(2017)\citenamefont {Chen},
		\citenamefont {Zhu}, \citenamefont {Yang},\ and\ \citenamefont
		{Wen}}]{HighlyAnisotropicSuperconductingchen2017}%
	\BibitemOpen
	\bibfield  {author} {\bibinfo {author} {\bibfnamefont {G.-Y.}\ \bibnamefont
			{Chen}}, \bibinfo {author} {\bibfnamefont {X.}~\bibnamefont {Zhu}}, \bibinfo
		{author} {\bibfnamefont {H.}~\bibnamefont {Yang}},\ and\ \bibinfo {author}
		{\bibfnamefont {H.-H.}\ \bibnamefont {Wen}},\ }\bibfield  {title} {\bibinfo
		{title} {Highly anisotropic superconducting gaps and possible evidence of
			antiferromagnetic order in {{FeSe}} single crystals},\ }\href
	{https://doi.org/10.1103/PhysRevB.96.064524} {\bibfield  {journal} {\bibinfo
			{journal} {Phys. Rev. B}\ }\textbf {\bibinfo {volume} {96}},\ \bibinfo
		{pages} {064524} (\bibinfo {year} {2017})}\BibitemShut {NoStop}%
	\bibitem [{\citenamefont {Jiao}\ \emph {et~al.}(2017)\citenamefont {Jiao},
		\citenamefont {Huang}, \citenamefont {R\"{o}\ss~ler}, \citenamefont {Koz},
		\citenamefont {R\"{o}\ss~ler}, \citenamefont {Schwarz},\ and\ \citenamefont
		{Wirth}}]{SuperconductingGapStructurejiao2017a}%
	\BibitemOpen
	\bibfield  {author} {\bibinfo {author} {\bibfnamefont {L.}~\bibnamefont
			{Jiao}}, \bibinfo {author} {\bibfnamefont {C.-L.}\ \bibnamefont {Huang}},
		\bibinfo {author} {\bibfnamefont {S.}~\bibnamefont {R\"{o}\ss~ler}}, \bibinfo
		{author} {\bibfnamefont {C.}~\bibnamefont {Koz}}, \bibinfo {author}
		{\bibfnamefont {U.~K.}\ \bibnamefont {R\"{o}\ss~ler}}, \bibinfo {author}
		{\bibfnamefont {U.}~\bibnamefont {Schwarz}},\ and\ \bibinfo {author}
		{\bibfnamefont {S.}~\bibnamefont {Wirth}},\ }\bibfield  {title} {\bibinfo
		{title} {Superconducting gap structure of {{FeSe}}},\ }\href
	{https://doi.org/10.1038/srep44024} {\bibfield  {journal} {\bibinfo
			{journal} {Scientific Reports}\ }\textbf {\bibinfo {volume} {7}},\ \bibinfo
		{pages} {44024} (\bibinfo {year} {2017})}\BibitemShut {NoStop}%
	\bibitem [{\citenamefont {Sprau}\ \emph {et~al.}(2017)\citenamefont {Sprau},
		\citenamefont {Kostin}, \citenamefont {Kreisel}, \citenamefont {B\"{o}hmer},
		\citenamefont {Taufour}, \citenamefont {Canfield}, \citenamefont {Mukherjee},
		\citenamefont {Hirschfeld}, \citenamefont {Andersen},\ and\ \citenamefont
		{Davis}}]{DiscoveryOrbitalselectiveCoopersprau2017}%
	\BibitemOpen
	\bibfield  {author} {\bibinfo {author} {\bibfnamefont {P.~O.}\ \bibnamefont
			{Sprau}}, \bibinfo {author} {\bibfnamefont {A.}~\bibnamefont {Kostin}},
		\bibinfo {author} {\bibfnamefont {A.}~\bibnamefont {Kreisel}}, \bibinfo
		{author} {\bibfnamefont {A.~E.}\ \bibnamefont {B\"{o}hmer}}, \bibinfo
		{author} {\bibfnamefont {V.}~\bibnamefont {Taufour}}, \bibinfo {author}
		{\bibfnamefont {P.~C.}\ \bibnamefont {Canfield}}, \bibinfo {author}
		{\bibfnamefont {S.}~\bibnamefont {Mukherjee}}, \bibinfo {author}
		{\bibfnamefont {P.~J.}\ \bibnamefont {Hirschfeld}}, \bibinfo {author}
		{\bibfnamefont {B.~M.}\ \bibnamefont {Andersen}},\ and\ \bibinfo {author}
		{\bibfnamefont {J.~C.~S.}\ \bibnamefont {Davis}},\ }\bibfield  {title}
	{\bibinfo {title} {Discovery of orbital-selective {{Cooper}} pairing in
			{{FeSe}}},\ }\href {https://doi.org/10.1126/science.aal1575} {\bibfield
		{journal} {\bibinfo  {journal} {Science}\ }\textbf {\bibinfo {volume}
			{357}},\ \bibinfo {pages} {75} (\bibinfo {year} {2017})}\BibitemShut
	{NoStop}%
	\bibitem [{\citenamefont {Hashimoto}\ \emph {et~al.}(2018)\citenamefont
		{Hashimoto}, \citenamefont {Ota}, \citenamefont {Yamamoto}, \citenamefont
		{Suzuki}, \citenamefont {Shimojima}, \citenamefont {Watanabe}, \citenamefont
		{Chen}, \citenamefont {Kasahara}, \citenamefont {Matsuda}, \citenamefont
		{Shibauchi}, \citenamefont {Okazaki},\ and\ \citenamefont
		{Shin}}]{SuperconductingGapAnisotropyhashimoto2018}%
	\BibitemOpen
	\bibfield  {author} {\bibinfo {author} {\bibfnamefont {T.}~\bibnamefont
			{Hashimoto}}, \bibinfo {author} {\bibfnamefont {Y.}~\bibnamefont {Ota}},
		\bibinfo {author} {\bibfnamefont {H.~Q.}\ \bibnamefont {Yamamoto}}, \bibinfo
		{author} {\bibfnamefont {Y.}~\bibnamefont {Suzuki}}, \bibinfo {author}
		{\bibfnamefont {T.}~\bibnamefont {Shimojima}}, \bibinfo {author}
		{\bibfnamefont {S.}~\bibnamefont {Watanabe}}, \bibinfo {author}
		{\bibfnamefont {C.}~\bibnamefont {Chen}}, \bibinfo {author} {\bibfnamefont
			{S.}~\bibnamefont {Kasahara}}, \bibinfo {author} {\bibfnamefont
			{Y.}~\bibnamefont {Matsuda}}, \bibinfo {author} {\bibfnamefont
			{T.}~\bibnamefont {Shibauchi}}, \bibinfo {author} {\bibfnamefont
			{K.}~\bibnamefont {Okazaki}},\ and\ \bibinfo {author} {\bibfnamefont
			{S.}~\bibnamefont {Shin}},\ }\bibfield  {title} {\bibinfo {title}
		{Superconducting gap anisotropy sensitive to nematic domains in {{FeSe}}},\
	}\href {https://doi.org/10.1038/s41467-017-02739-y} {\bibfield  {journal}
		{\bibinfo  {journal} {Nature Communications}\ }\textbf {\bibinfo {volume}
			{9}},\ \bibinfo {pages} {282} (\bibinfo {year} {2018})}\BibitemShut {NoStop}%
	\bibitem [{\citenamefont {Qian}\ \emph {et~al.}(2011)\citenamefont {Qian},
		\citenamefont {Wang}, \citenamefont {Jin}, \citenamefont {Zhang},
		\citenamefont {Richard}, \citenamefont {Xu}, \citenamefont {Dai},
		\citenamefont {Fang}, \citenamefont {Guo}, \citenamefont {Chen},\ and\
		\citenamefont {Ding}}]{Qian2011}%
	\BibitemOpen
	\bibfield  {author} {\bibinfo {author} {\bibfnamefont {T.}~\bibnamefont
			{Qian}}, \bibinfo {author} {\bibfnamefont {X.-P.}\ \bibnamefont {Wang}},
		\bibinfo {author} {\bibfnamefont {W.-C.}\ \bibnamefont {Jin}}, \bibinfo
		{author} {\bibfnamefont {P.}~\bibnamefont {Zhang}}, \bibinfo {author}
		{\bibfnamefont {P.}~\bibnamefont {Richard}}, \bibinfo {author} {\bibfnamefont
			{G.}~\bibnamefont {Xu}}, \bibinfo {author} {\bibfnamefont {X.}~\bibnamefont
			{Dai}}, \bibinfo {author} {\bibfnamefont {Z.}~\bibnamefont {Fang}}, \bibinfo
		{author} {\bibfnamefont {J.-G.}\ \bibnamefont {Guo}}, \bibinfo {author}
		{\bibfnamefont {X.-L.}\ \bibnamefont {Chen}},\ and\ \bibinfo {author}
		{\bibfnamefont {H.}~\bibnamefont {Ding}},\ }\bibfield  {title} {\bibinfo
		{title} {Absence of a holelike {F}ermi surface for the iron-based
			{${\mathrm{K}}_{0.8}{\mathrm{Fe}}_{1.7}{\mathrm{Se}}_{2}$} superconductor
			revealed by angle-resolved photoemission spectroscopy},\ }\href
	{https://doi.org/10.1103/PhysRevLett.106.187001} {\bibfield  {journal}
		{\bibinfo  {journal} {Phys. Rev. Lett.}\ }\textbf {\bibinfo {volume} {106}},\
		\bibinfo {pages} {187001} (\bibinfo {year} {2011})}\BibitemShut {NoStop}%
	\bibitem [{\citenamefont {Liu}\ \emph {et~al.}(2012)\citenamefont {Liu},
		\citenamefont {Zhang}, \citenamefont {Mou}, \citenamefont {He}, \citenamefont
		{Ou}, \citenamefont {Wang}, \citenamefont {Li}, \citenamefont {Wang},
		\citenamefont {Zhao}, \citenamefont {He}, \citenamefont {Peng}, \citenamefont
		{Liu}, \citenamefont {Chen}, \citenamefont {Yu}, \citenamefont {Liu},
		\citenamefont {Dong}, \citenamefont {Zhang}, \citenamefont {Chen},
		\citenamefont {Xu}, \citenamefont {Hu}, \citenamefont {Chen}, \citenamefont
		{Ma}, \citenamefont {Xue},\ and\ \citenamefont
		{Zhou}}]{ElectronicOriginHightemperatureliu2012}%
	\BibitemOpen
	\bibfield  {author} {\bibinfo {author} {\bibfnamefont {D.}~\bibnamefont
			{Liu}}, \bibinfo {author} {\bibfnamefont {W.}~\bibnamefont {Zhang}}, \bibinfo
		{author} {\bibfnamefont {D.}~\bibnamefont {Mou}}, \bibinfo {author}
		{\bibfnamefont {J.}~\bibnamefont {He}}, \bibinfo {author} {\bibfnamefont
			{Y.-B.}\ \bibnamefont {Ou}}, \bibinfo {author} {\bibfnamefont {Q.-Y.}\
			\bibnamefont {Wang}}, \bibinfo {author} {\bibfnamefont {Z.}~\bibnamefont
			{Li}}, \bibinfo {author} {\bibfnamefont {L.}~\bibnamefont {Wang}}, \bibinfo
		{author} {\bibfnamefont {L.}~\bibnamefont {Zhao}}, \bibinfo {author}
		{\bibfnamefont {S.}~\bibnamefont {He}}, \bibinfo {author} {\bibfnamefont
			{Y.}~\bibnamefont {Peng}}, \bibinfo {author} {\bibfnamefont {X.}~\bibnamefont
			{Liu}}, \bibinfo {author} {\bibfnamefont {C.}~\bibnamefont {Chen}}, \bibinfo
		{author} {\bibfnamefont {L.}~\bibnamefont {Yu}}, \bibinfo {author}
		{\bibfnamefont {G.}~\bibnamefont {Liu}}, \bibinfo {author} {\bibfnamefont
			{X.}~\bibnamefont {Dong}}, \bibinfo {author} {\bibfnamefont {J.}~\bibnamefont
			{Zhang}}, \bibinfo {author} {\bibfnamefont {C.}~\bibnamefont {Chen}},
		\bibinfo {author} {\bibfnamefont {Z.}~\bibnamefont {Xu}}, \bibinfo {author}
		{\bibfnamefont {J.}~\bibnamefont {Hu}}, \bibinfo {author} {\bibfnamefont
			{X.}~\bibnamefont {Chen}}, \bibinfo {author} {\bibfnamefont {X.}~\bibnamefont
			{Ma}}, \bibinfo {author} {\bibfnamefont {Q.}~\bibnamefont {Xue}},\ and\
		\bibinfo {author} {\bibfnamefont {X.}~\bibnamefont {Zhou}},\ }\bibfield
	{title} {\bibinfo {title} {Electronic origin of high-temperature
			superconductivity in single-layer {{FeSe}} superconductor},\ }\href
	{https://doi.org/10.1038/ncomms1946} {\bibfield  {journal} {\bibinfo
			{journal} {Nature Communications}\ }\textbf {\bibinfo {volume} {3}},\
		\bibinfo {pages} {931} (\bibinfo {year} {2012})}\BibitemShut {NoStop}%
	\bibitem [{\citenamefont {Zhao}\ \emph {et~al.}(2016)\citenamefont {Zhao},
		\citenamefont {Liang}, \citenamefont {Yuan}, \citenamefont {Hu},
		\citenamefont {Liu}, \citenamefont {Huang}, \citenamefont {He}, \citenamefont
		{Shen}, \citenamefont {Xu}, \citenamefont {Liu} \emph
		{et~al.}}]{CommonElectronicOriginzhao}%
	\BibitemOpen
	\bibfield  {author} {\bibinfo {author} {\bibfnamefont {L.}~\bibnamefont
			{Zhao}}, \bibinfo {author} {\bibfnamefont {A.}~\bibnamefont {Liang}},
		\bibinfo {author} {\bibfnamefont {D.}~\bibnamefont {Yuan}}, \bibinfo {author}
		{\bibfnamefont {Y.}~\bibnamefont {Hu}}, \bibinfo {author} {\bibfnamefont
			{D.}~\bibnamefont {Liu}}, \bibinfo {author} {\bibfnamefont {J.}~\bibnamefont
			{Huang}}, \bibinfo {author} {\bibfnamefont {S.}~\bibnamefont {He}}, \bibinfo
		{author} {\bibfnamefont {B.}~\bibnamefont {Shen}}, \bibinfo {author}
		{\bibfnamefont {Y.}~\bibnamefont {Xu}}, \bibinfo {author} {\bibfnamefont
			{X.}~\bibnamefont {Liu}}, \emph {et~al.},\ }\bibfield  {title} {\bibinfo
		{title} {Common electronic origin of superconductivity in
			({{Li}},{{Fe}}){{OHFeSe}} bulk superconductor and single-layer
			{{FeSe}}/{{SrTiO$_3$}} films},\ }\href {https://doi.org/10.1038/ncomms10608}
	{\bibfield  {journal} {\bibinfo  {journal} {Nature Communications}\ }\textbf
		{\bibinfo {volume} {7}},\ \bibinfo {pages} {10608} (\bibinfo {year}
		{2016})}\BibitemShut {NoStop}%
	\bibitem [{\citenamefont {Niu}\ \emph {et~al.}(2015)\citenamefont {Niu},
		\citenamefont {Peng}, \citenamefont {Xu}, \citenamefont {Yan}, \citenamefont
		{Jiang}, \citenamefont {Xu}, \citenamefont {Yu}, \citenamefont {Song},
		\citenamefont {Huang}, \citenamefont {Wang}, \citenamefont {Xie},
		\citenamefont {Lu}, \citenamefont {Wang}, \citenamefont {Chen}, \citenamefont
		{Sun},\ and\ \citenamefont {Feng}}]{SurfaceElectronicStructureniu2015}%
	\BibitemOpen
	\bibfield  {author} {\bibinfo {author} {\bibfnamefont {X.~H.}\ \bibnamefont
			{Niu}}, \bibinfo {author} {\bibfnamefont {R.}~\bibnamefont {Peng}}, \bibinfo
		{author} {\bibfnamefont {H.~C.}\ \bibnamefont {Xu}}, \bibinfo {author}
		{\bibfnamefont {Y.~J.}\ \bibnamefont {Yan}}, \bibinfo {author} {\bibfnamefont
			{J.}~\bibnamefont {Jiang}}, \bibinfo {author} {\bibfnamefont {D.~F.}\
			\bibnamefont {Xu}}, \bibinfo {author} {\bibfnamefont {T.~L.}\ \bibnamefont
			{Yu}}, \bibinfo {author} {\bibfnamefont {Q.}~\bibnamefont {Song}}, \bibinfo
		{author} {\bibfnamefont {Z.~C.}\ \bibnamefont {Huang}}, \bibinfo {author}
		{\bibfnamefont {Y.~X.}\ \bibnamefont {Wang}}, \bibinfo {author}
		{\bibfnamefont {B.~P.}\ \bibnamefont {Xie}}, \bibinfo {author} {\bibfnamefont
			{X.~F.}\ \bibnamefont {Lu}}, \bibinfo {author} {\bibfnamefont {N.~Z.}\
			\bibnamefont {Wang}}, \bibinfo {author} {\bibfnamefont {X.~H.}\ \bibnamefont
			{Chen}}, \bibinfo {author} {\bibfnamefont {Z.}~\bibnamefont {Sun}},\ and\
		\bibinfo {author} {\bibfnamefont {D.~L.}\ \bibnamefont {Feng}},\ }\bibfield
	{title} {\bibinfo {title} {Surface electronic structure and isotropic
			superconducting gap in ({{Li$_{0.8}$Fe$_{0.2}$)OHFeSe}}},\ }\href
	{https://doi.org/10.1103/PhysRevB.92.060504} {\bibfield  {journal} {\bibinfo
			{journal} {Phys. Rev. B}\ }\textbf {\bibinfo {volume} {92}},\ \bibinfo
		{pages} {060504(R)} (\bibinfo {year} {2015})}\BibitemShut {NoStop}%
	\bibitem [{\citenamefont {Zhang}\ \emph {et~al.}(2016)\citenamefont {Zhang},
		\citenamefont {Lee}, \citenamefont {Moore}, \citenamefont {Li}, \citenamefont
		{Yi}, \citenamefont {Hashimoto}, \citenamefont {Lu}, \citenamefont
		{Devereaux}, \citenamefont {Lee},\ and\ \citenamefont
		{Shen}}]{SuperconductingGapAnisotropyzhang2016}%
	\BibitemOpen
	\bibfield  {author} {\bibinfo {author} {\bibfnamefont {Y.}~\bibnamefont
			{Zhang}}, \bibinfo {author} {\bibfnamefont {J.~J.}\ \bibnamefont {Lee}},
		\bibinfo {author} {\bibfnamefont {R.~G.}\ \bibnamefont {Moore}}, \bibinfo
		{author} {\bibfnamefont {W.}~\bibnamefont {Li}}, \bibinfo {author}
		{\bibfnamefont {M.}~\bibnamefont {Yi}}, \bibinfo {author} {\bibfnamefont
			{M.}~\bibnamefont {Hashimoto}}, \bibinfo {author} {\bibfnamefont {D.~H.}\
			\bibnamefont {Lu}}, \bibinfo {author} {\bibfnamefont {T.~P.}\ \bibnamefont
			{Devereaux}}, \bibinfo {author} {\bibfnamefont {D.-H.}\ \bibnamefont {Lee}},\
		and\ \bibinfo {author} {\bibfnamefont {Z.-X.}\ \bibnamefont {Shen}},\
	}\bibfield  {title} {\bibinfo {title} {Superconducting {{Gap Anisotropy}} in
			{{Monolayer FeSe Thin Film}}},\ }\href
	{https://doi.org/10.1103/PhysRevLett.117.117001} {\bibfield  {journal}
		{\bibinfo  {journal} {Phys. Rev. Lett.}\ }\textbf {\bibinfo {volume} {117}},\
		\bibinfo {pages} {117001} (\bibinfo {year} {2016})}\BibitemShut {NoStop}%
	\bibitem [{\citenamefont {Du}\ \emph {et~al.}(2016)\citenamefont {Du},
		\citenamefont {Yang}, \citenamefont {Lin}, \citenamefont {Fang},
		\citenamefont {Du}, \citenamefont {Xing}, \citenamefont {Yang}, \citenamefont
		{Zhu},\ and\ \citenamefont {Wen}}]{ScrutinizingDoubleSuperconductingdu2016}%
	\BibitemOpen
	\bibfield  {author} {\bibinfo {author} {\bibfnamefont {Z.}~\bibnamefont
			{Du}}, \bibinfo {author} {\bibfnamefont {X.}~\bibnamefont {Yang}}, \bibinfo
		{author} {\bibfnamefont {H.}~\bibnamefont {Lin}}, \bibinfo {author}
		{\bibfnamefont {D.}~\bibnamefont {Fang}}, \bibinfo {author} {\bibfnamefont
			{G.}~\bibnamefont {Du}}, \bibinfo {author} {\bibfnamefont {J.}~\bibnamefont
			{Xing}}, \bibinfo {author} {\bibfnamefont {H.}~\bibnamefont {Yang}}, \bibinfo
		{author} {\bibfnamefont {X.}~\bibnamefont {Zhu}},\ and\ \bibinfo {author}
		{\bibfnamefont {H.-H.}\ \bibnamefont {Wen}},\ }\bibfield  {title} {\bibinfo
		{title} {Scrutinizing the double superconducting gaps and strong coupling
			pairing in ({{Li$_{1-x}$Fe$_x$)OHFeSe}}},\ }\href
	{https://doi.org/10.1038/ncomms10565} {\bibfield  {journal} {\bibinfo
			{journal} {Nature Communications}\ }\textbf {\bibinfo {volume} {7}},\
		\bibinfo {pages} {10565} (\bibinfo {year} {2016})}\BibitemShut {NoStop}%
	\bibitem [{\citenamefont {Yan}\ \emph {et~al.}(2016)\citenamefont {Yan},
		\citenamefont {Zhang}, \citenamefont {Ren}, \citenamefont {Liu},
		\citenamefont {Lu}, \citenamefont {Wang}, \citenamefont {Niu}, \citenamefont
		{Fan}, \citenamefont {Miao}, \citenamefont {Tao}, \citenamefont {Xie},
		\citenamefont {Chen}, \citenamefont {Zhang},\ and\ \citenamefont
		{Feng}}]{SurfaceElectronicStructureyan2016}%
	\BibitemOpen
	\bibfield  {author} {\bibinfo {author} {\bibfnamefont {Y.~J.}\ \bibnamefont
			{Yan}}, \bibinfo {author} {\bibfnamefont {W.~H.}\ \bibnamefont {Zhang}},
		\bibinfo {author} {\bibfnamefont {M.~Q.}\ \bibnamefont {Ren}}, \bibinfo
		{author} {\bibfnamefont {X.}~\bibnamefont {Liu}}, \bibinfo {author}
		{\bibfnamefont {X.~F.}\ \bibnamefont {Lu}}, \bibinfo {author} {\bibfnamefont
			{N.~Z.}\ \bibnamefont {Wang}}, \bibinfo {author} {\bibfnamefont {X.~H.}\
			\bibnamefont {Niu}}, \bibinfo {author} {\bibfnamefont {Q.}~\bibnamefont
			{Fan}}, \bibinfo {author} {\bibfnamefont {J.}~\bibnamefont {Miao}}, \bibinfo
		{author} {\bibfnamefont {R.}~\bibnamefont {Tao}}, \bibinfo {author}
		{\bibfnamefont {B.~P.}\ \bibnamefont {Xie}}, \bibinfo {author} {\bibfnamefont
			{X.~H.}\ \bibnamefont {Chen}}, \bibinfo {author} {\bibfnamefont
			{T.}~\bibnamefont {Zhang}},\ and\ \bibinfo {author} {\bibfnamefont {D.~L.}\
			\bibnamefont {Feng}},\ }\bibfield  {title} {\bibinfo {title} {Surface
			electronic structure and evidence of plain s-wave superconductivity in
			({{Li$_{0.8}$Fe$_{0.2}$)OHFeSe}}},\ }\href
	{https://doi.org/10.1103/PhysRevB.94.134502} {\bibfield  {journal} {\bibinfo
			{journal} {Phys. Rev. B}\ }\textbf {\bibinfo {volume} {94}},\ \bibinfo
		{pages} {134502} (\bibinfo {year} {2016})}\BibitemShut {NoStop}%
	\bibitem [{\citenamefont {Khasanov}\ \emph {et~al.}(2016)\citenamefont
		{Khasanov}, \citenamefont {Zhou}, \citenamefont {Amato}, \citenamefont
		{Guguchia}, \citenamefont {Morenzoni}, \citenamefont {Dong}, \citenamefont
		{Zhang},\ and\ \citenamefont
		{Zhao}}]{khasanov2016ProximityinducedSuperconductivityInsulating}%
	\BibitemOpen
	\bibfield  {author} {\bibinfo {author} {\bibfnamefont {R.}~\bibnamefont
			{Khasanov}}, \bibinfo {author} {\bibfnamefont {H.}~\bibnamefont {Zhou}},
		\bibinfo {author} {\bibfnamefont {A.}~\bibnamefont {Amato}}, \bibinfo
		{author} {\bibfnamefont {Z.}~\bibnamefont {Guguchia}}, \bibinfo {author}
		{\bibfnamefont {E.}~\bibnamefont {Morenzoni}}, \bibinfo {author}
		{\bibfnamefont {X.}~\bibnamefont {Dong}}, \bibinfo {author} {\bibfnamefont
			{G.}~\bibnamefont {Zhang}},\ and\ \bibinfo {author} {\bibfnamefont
			{Z.}~\bibnamefont {Zhao}},\ }\bibfield  {title} {\bibinfo {title}
		{Proximity-induced superconductivity within the insulating
			({{Li$_{0.84}$Fe$_{0.16}$)OH}} layers in
			({{Li$_{0.84}$Fe$_{0.16}$)OHFe$_{0.98}$Se}}},\ }\href
	{https://doi.org/10.1103/PhysRevB.93.224512} {\bibfield  {journal} {\bibinfo
			{journal} {Phys. Rev. B}\ }\textbf {\bibinfo {volume} {93}},\ \bibinfo
		{pages} {224512} (\bibinfo {year} {2016})}\BibitemShut {NoStop}%
	\bibitem [{\citenamefont {Smidman}\ \emph {et~al.}(2017)\citenamefont
		{Smidman}, \citenamefont {Pang}, \citenamefont {Zhou}, \citenamefont {Wang},
		\citenamefont {Xie}, \citenamefont {Weng}, \citenamefont {Chen},
		\citenamefont {Dong}, \citenamefont {Chen}, \citenamefont {Zhao},\ and\
		\citenamefont {Yuan}}]{Smidman2017}%
	\BibitemOpen
	\bibfield  {author} {\bibinfo {author} {\bibfnamefont {M.}~\bibnamefont
			{Smidman}}, \bibinfo {author} {\bibfnamefont {G.~M.}\ \bibnamefont {Pang}},
		\bibinfo {author} {\bibfnamefont {H.~X.}\ \bibnamefont {Zhou}}, \bibinfo
		{author} {\bibfnamefont {N.~Z.}\ \bibnamefont {Wang}}, \bibinfo {author}
		{\bibfnamefont {W.}~\bibnamefont {Xie}}, \bibinfo {author} {\bibfnamefont
			{Z.~F.}\ \bibnamefont {Weng}}, \bibinfo {author} {\bibfnamefont
			{Y.}~\bibnamefont {Chen}}, \bibinfo {author} {\bibfnamefont {X.~L.}\
			\bibnamefont {Dong}}, \bibinfo {author} {\bibfnamefont {X.~H.}\ \bibnamefont
			{Chen}}, \bibinfo {author} {\bibfnamefont {Z.~X.}\ \bibnamefont {Zhao}},\
		and\ \bibinfo {author} {\bibfnamefont {H.~Q.}\ \bibnamefont {Yuan}},\
	}\bibfield  {title} {\bibinfo {title} {Probing the superconducting gap
			structure of
			{$({\mathrm{Li}}_{1\ensuremath{-}x}{\mathrm{Fe}}_{x})\mathrm{OHFeSe}$}},\
	}\href {https://doi.org/10.1103/PhysRevB.96.014504} {\bibfield  {journal}
		{\bibinfo  {journal} {Phys. Rev. B}\ }\textbf {\bibinfo {volume} {96}},\
		\bibinfo {pages} {014504} (\bibinfo {year} {2017})}\BibitemShut {NoStop}%
	\bibitem [{\citenamefont {Du}\ \emph {et~al.}(2018)\citenamefont {Du},
		\citenamefont {Yang}, \citenamefont {Altenfeld}, \citenamefont {Gu},
		\citenamefont {Yang}, \citenamefont {Eremin}, \citenamefont {Hirschfeld},
		\citenamefont {Mazin}, \citenamefont {Lin}, \citenamefont {Zhu},\ and\
		\citenamefont {Wen}}]{Du2018}%
	\BibitemOpen
	\bibfield  {author} {\bibinfo {author} {\bibfnamefont {Z.}~\bibnamefont
			{Du}}, \bibinfo {author} {\bibfnamefont {X.}~\bibnamefont {Yang}}, \bibinfo
		{author} {\bibfnamefont {D.}~\bibnamefont {Altenfeld}}, \bibinfo {author}
		{\bibfnamefont {Q.}~\bibnamefont {Gu}}, \bibinfo {author} {\bibfnamefont
			{H.}~\bibnamefont {Yang}}, \bibinfo {author} {\bibfnamefont {I.}~\bibnamefont
			{Eremin}}, \bibinfo {author} {\bibfnamefont {P.~J.}\ \bibnamefont
			{Hirschfeld}}, \bibinfo {author} {\bibfnamefont {I.~I.}\ \bibnamefont
			{Mazin}}, \bibinfo {author} {\bibfnamefont {H.}~\bibnamefont {Lin}}, \bibinfo
		{author} {\bibfnamefont {X.}~\bibnamefont {Zhu}},\ and\ \bibinfo {author}
		{\bibfnamefont {H.-H.}\ \bibnamefont {Wen}},\ }\bibfield  {title} {\bibinfo
		{title} {Sign reversal of the order parameter in
			({Li}$_{1-x}${Fe}$_x$){OHFe}$_{1-y}${Zn}$_y${Se}},\ }\href
	{https://doi.org/10.1038/nphys4299} {\bibfield  {journal} {\bibinfo
			{journal} {Nature Physics}\ }\textbf {\bibinfo {volume} {14}},\ \bibinfo
		{pages} {134} (\bibinfo {year} {2018})}\BibitemShut {NoStop}%
	\bibitem [{\citenamefont {Shi}\ \emph {et~al.}(2018{\natexlab{a}})\citenamefont
		{Shi}, \citenamefont {Wang}, \citenamefont {Lei}, \citenamefont {Ying},
		\citenamefont {Zhu}, \citenamefont {Sun}, \citenamefont {Cui}, \citenamefont
		{Meng}, \citenamefont {Shang}, \citenamefont {Ma},\ and\ \citenamefont
		{Chen}}]{shi2018FeSebasedSuperconductorsSuperconducting}%
	\BibitemOpen
	\bibfield  {author} {\bibinfo {author} {\bibfnamefont {M.~Z.}\ \bibnamefont
			{Shi}}, \bibinfo {author} {\bibfnamefont {N.~Z.}\ \bibnamefont {Wang}},
		\bibinfo {author} {\bibfnamefont {B.}~\bibnamefont {Lei}}, \bibinfo {author}
		{\bibfnamefont {J.~J.}\ \bibnamefont {Ying}}, \bibinfo {author}
		{\bibfnamefont {C.~S.}\ \bibnamefont {Zhu}}, \bibinfo {author} {\bibfnamefont
			{Z.~L.}\ \bibnamefont {Sun}}, \bibinfo {author} {\bibfnamefont {J.~H.}\
			\bibnamefont {Cui}}, \bibinfo {author} {\bibfnamefont {F.~B.}\ \bibnamefont
			{Meng}}, \bibinfo {author} {\bibfnamefont {C.}~\bibnamefont {Shang}},
		\bibinfo {author} {\bibfnamefont {L.~K.}\ \bibnamefont {Ma}},\ and\ \bibinfo
		{author} {\bibfnamefont {X.~H.}\ \bibnamefont {Chen}},\ }\bibfield  {title}
	{\bibinfo {title} {{{FeSe-based}} superconductors with a superconducting
			transition temperature of 50 {{K}}},\ }\href
	{https://doi.org/10.1088/1367-2630/aaf312} {\bibfield  {journal} {\bibinfo
			{journal} {New Journal of Physics}\ }\textbf {\bibinfo {volume} {20}},\
		\bibinfo {pages} {123007} (\bibinfo {year} {2018}{\natexlab{a}})}\BibitemShut
	{NoStop}%
	\bibitem [{\citenamefont {Shi}\ \emph {et~al.}(2018{\natexlab{b}})\citenamefont
		{Shi}, \citenamefont {Wang}, \citenamefont {Lei}, \citenamefont {Shang},
		\citenamefont {Meng}, \citenamefont {Ma}, \citenamefont {Zhang},
		\citenamefont {Kuang},\ and\ \citenamefont
		{Chen}}]{OrganicionintercalatedFeSebasedSuperconductorsshi2018}%
	\BibitemOpen
	\bibfield  {author} {\bibinfo {author} {\bibfnamefont {M.~Z.}\ \bibnamefont
			{Shi}}, \bibinfo {author} {\bibfnamefont {N.~Z.}\ \bibnamefont {Wang}},
		\bibinfo {author} {\bibfnamefont {B.}~\bibnamefont {Lei}}, \bibinfo {author}
		{\bibfnamefont {C.}~\bibnamefont {Shang}}, \bibinfo {author} {\bibfnamefont
			{F.~B.}\ \bibnamefont {Meng}}, \bibinfo {author} {\bibfnamefont {L.~K.}\
			\bibnamefont {Ma}}, \bibinfo {author} {\bibfnamefont {F.~X.}\ \bibnamefont
			{Zhang}}, \bibinfo {author} {\bibfnamefont {D.~Z.}\ \bibnamefont {Kuang}},\
		and\ \bibinfo {author} {\bibfnamefont {X.~H.}\ \bibnamefont {Chen}},\
	}\bibfield  {title} {\bibinfo {title} {Organic-ion-intercalated
			{{FeSe-based}} superconductors},\ }\href
	{https://doi.org/10.1103/PhysRevMaterials.2.074801} {\bibfield  {journal}
		{\bibinfo  {journal} {Physical Review Materials}\ }\textbf {\bibinfo {volume}
			{2}},\ \bibinfo {pages} {074801} (\bibinfo {year}
		{2018}{\natexlab{b}})}\BibitemShut {NoStop}%
	\bibitem [{\citenamefont {Kang}\ \emph {et~al.}(2020)\citenamefont {Kang},
		\citenamefont {Shi}, \citenamefont {Li}, \citenamefont {Wang}, \citenamefont
		{Zhang}, \citenamefont {Zhao}, \citenamefont {Li}, \citenamefont {Song},
		\citenamefont {Zheng}, \citenamefont {Nie}, \citenamefont {Wu},\ and\
		\citenamefont {Chen}}]{kang2020PreformedCooperPairs}%
	\BibitemOpen
	\bibfield  {author} {\bibinfo {author} {\bibfnamefont {B.~L.}\ \bibnamefont
			{Kang}}, \bibinfo {author} {\bibfnamefont {M.~Z.}\ \bibnamefont {Shi}},
		\bibinfo {author} {\bibfnamefont {S.~J.}\ \bibnamefont {Li}}, \bibinfo
		{author} {\bibfnamefont {H.~H.}\ \bibnamefont {Wang}}, \bibinfo {author}
		{\bibfnamefont {Q.}~\bibnamefont {Zhang}}, \bibinfo {author} {\bibfnamefont
			{D.}~\bibnamefont {Zhao}}, \bibinfo {author} {\bibfnamefont {J.}~\bibnamefont
			{Li}}, \bibinfo {author} {\bibfnamefont {D.~W.}\ \bibnamefont {Song}},
		\bibinfo {author} {\bibfnamefont {L.~X.}\ \bibnamefont {Zheng}}, \bibinfo
		{author} {\bibfnamefont {L.~P.}\ \bibnamefont {Nie}}, \bibinfo {author}
		{\bibfnamefont {T.}~\bibnamefont {Wu}},\ and\ \bibinfo {author}
		{\bibfnamefont {X.~H.}\ \bibnamefont {Chen}},\ }\bibfield  {title} {\bibinfo
		{title} {Preformed {{Cooper Pairs}} in {{Layered FeSe-Based
					Superconductors}}},\ }\href {https://doi.org/10.1103/PhysRevLett.125.097003}
	{\bibfield  {journal} {\bibinfo  {journal} {Phys. Rev. Lett.}\ }\textbf
		{\bibinfo {volume} {125}},\ \bibinfo {pages} {097003} (\bibinfo {year}
		{2020})}\BibitemShut {NoStop}%
	\bibitem [{\citenamefont {Pratt}(2000)}]{pratt2000WIMDAMuonData}%
	\BibitemOpen
	\bibfield  {author} {\bibinfo {author} {\bibfnamefont {F.~L.}\ \bibnamefont
			{Pratt}},\ }\bibfield  {title} {\bibinfo {title} {{{WIMDA}}: A muon data
			analysis program for the {{Windows PC}}},\ }\href
	{https://doi.org/10.1016/S0921-4526(00)00328-8} {\bibfield  {journal}
		{\bibinfo  {journal} {Physica B: Condensed Matter}\ }\textbf {\bibinfo
			{volume} {289--290}},\ \bibinfo {pages} {710} (\bibinfo {year}
		{2000})}\BibitemShut {NoStop}%
	\bibitem [{\citenamefont
		{Brandt}(2003)}]{brandt2003PropertiesIdealGinzburgLandau}%
	\BibitemOpen
	\bibfield  {author} {\bibinfo {author} {\bibfnamefont {E.~H.}\ \bibnamefont
			{Brandt}},\ }\bibfield  {title} {\bibinfo {title} {Properties of the ideal
			{{Ginzburg-Landau}} vortex lattice},\ }\href
	{https://doi.org/10.1103/PhysRevB.68.054506} {\bibfield  {journal} {\bibinfo
			{journal} {Phys. Rev. B}\ }\textbf {\bibinfo {volume} {68}},\ \bibinfo
		{pages} {054506} (\bibinfo {year} {2003})}\BibitemShut {NoStop}%
	\bibitem [{\citenamefont {Tinkham}(2004)}]{tinkham2004introduction}%
	\BibitemOpen
	\bibfield  {author} {\bibinfo {author} {\bibfnamefont {M.}~\bibnamefont
			{Tinkham}},\ }\href@noop {} {\emph {\bibinfo {title} {Introduction to
				superconductivity\textnormal{, 2nd ed.}}}}\ (\bibinfo  {publisher} {Dover
			Publications, Mineola, NY},\ \bibinfo
	{year} {2004})\BibitemShut {NoStop}%
	\bibitem [{\citenamefont {Song}\ \emph {et~al.}(2011)\citenamefont {Song},
		\citenamefont {Wang}, \citenamefont {Cheng}, \citenamefont {Jiang},
		\citenamefont {Li}, \citenamefont {Zhang}, \citenamefont {Li}, \citenamefont
		{He}, \citenamefont {Wang}, \citenamefont {Jia}, \citenamefont {Hung},
		\citenamefont {Wu}, \citenamefont {Ma}, \citenamefont {Chen},\ and\
		\citenamefont {Xue}}]{science1202226}%
	\BibitemOpen
	\bibfield  {author} {\bibinfo {author} {\bibfnamefont {C.-L.}\ \bibnamefont
			{Song}}, \bibinfo {author} {\bibfnamefont {Y.-L.}\ \bibnamefont {Wang}},
		\bibinfo {author} {\bibfnamefont {P.}~\bibnamefont {Cheng}}, \bibinfo
		{author} {\bibfnamefont {Y.-P.}\ \bibnamefont {Jiang}}, \bibinfo {author}
		{\bibfnamefont {W.}~\bibnamefont {Li}}, \bibinfo {author} {\bibfnamefont
			{T.}~\bibnamefont {Zhang}}, \bibinfo {author} {\bibfnamefont
			{Z.}~\bibnamefont {Li}}, \bibinfo {author} {\bibfnamefont {K.}~\bibnamefont
			{He}}, \bibinfo {author} {\bibfnamefont {L.}~\bibnamefont {Wang}}, \bibinfo
		{author} {\bibfnamefont {J.-F.}\ \bibnamefont {Jia}}, \bibinfo {author}
		{\bibfnamefont {H.-H.}\ \bibnamefont {Hung}}, \bibinfo {author}
		{\bibfnamefont {C.}~\bibnamefont {Wu}}, \bibinfo {author} {\bibfnamefont
			{X.}~\bibnamefont {Ma}}, \bibinfo {author} {\bibfnamefont {X.}~\bibnamefont
			{Chen}},\ and\ \bibinfo {author} {\bibfnamefont {Q.-K.}\ \bibnamefont
			{Xue}},\ }\bibfield  {title} {\bibinfo {title} {{Direct Observation of Nodes
				and Twofold Symmetry in FeSe Superconductor}},\ }\href
	{https://doi.org/10.1126/science.1202226} {\bibfield  {journal} {\bibinfo
			{journal} {Science}\ }\textbf {\bibinfo {volume} {332}},\ \bibinfo {pages}
		{1410} (\bibinfo {year} {2011})}\BibitemShut {NoStop}%
	\bibitem [{\citenamefont {Fan}\ \emph {et~al.}(2015)\citenamefont {Fan},
		\citenamefont {Zhang}, \citenamefont {Liu}, \citenamefont {Yan},
		\citenamefont {Ren}, \citenamefont {Peng}, \citenamefont {Xu}, \citenamefont
		{Xie}, \citenamefont {Hu}, \citenamefont {Zhang} \emph
		{et~al.}}]{fan2015plain}%
	\BibitemOpen
	\bibfield  {author} {\bibinfo {author} {\bibfnamefont {Q.}~\bibnamefont
			{Fan}}, \bibinfo {author} {\bibfnamefont {W.}~\bibnamefont {Zhang}}, \bibinfo
		{author} {\bibfnamefont {X.}~\bibnamefont {Liu}}, \bibinfo {author}
		{\bibfnamefont {Y.}~\bibnamefont {Yan}}, \bibinfo {author} {\bibfnamefont
			{M.}~\bibnamefont {Ren}}, \bibinfo {author} {\bibfnamefont {R.}~\bibnamefont
			{Peng}}, \bibinfo {author} {\bibfnamefont {H.}~\bibnamefont {Xu}}, \bibinfo
		{author} {\bibfnamefont {B.}~\bibnamefont {Xie}}, \bibinfo {author}
		{\bibfnamefont {J.}~\bibnamefont {Hu}}, \bibinfo {author} {\bibfnamefont
			{T.}~\bibnamefont {Zhang}}, \emph {et~al.},\ }\bibfield  {title} {\bibinfo
		{title} {{Plain s-wave superconductivity in single-layer FeSe on SrTiO$_3$
				probed by scanning tunnelling microscopy}},\ }\href
	{https://doi.org/10.1038/nphys3450} {\bibfield  {journal} {\bibinfo
			{journal} {Nature Physics}\ }\textbf {\bibinfo {volume} {11}},\ \bibinfo
		{pages} {946} (\bibinfo {year} {2015})}\BibitemShut {NoStop}%
	\bibitem [{\citenamefont {Wang}\ \emph {et~al.}(2011)\citenamefont {Wang},
		\citenamefont {Qian}, \citenamefont {Richard}, \citenamefont {Zhang},
		\citenamefont {Dong}, \citenamefont {Wang}, \citenamefont {Dong},
		\citenamefont {Fang},\ and\ \citenamefont
		{Ding}}]{StrongNodelessPairingwang2011}%
	\BibitemOpen
	\bibfield  {author} {\bibinfo {author} {\bibfnamefont {X.-P.}\ \bibnamefont
			{Wang}}, \bibinfo {author} {\bibfnamefont {T.}~\bibnamefont {Qian}}, \bibinfo
		{author} {\bibfnamefont {P.}~\bibnamefont {Richard}}, \bibinfo {author}
		{\bibfnamefont {P.}~\bibnamefont {Zhang}}, \bibinfo {author} {\bibfnamefont
			{J.}~\bibnamefont {Dong}}, \bibinfo {author} {\bibfnamefont {H.-D.}\
			\bibnamefont {Wang}}, \bibinfo {author} {\bibfnamefont {C.-H.}\ \bibnamefont
			{Dong}}, \bibinfo {author} {\bibfnamefont {M.-H.}\ \bibnamefont {Fang}},\
		and\ \bibinfo {author} {\bibfnamefont {H.}~\bibnamefont {Ding}},\ }\bibfield
	{title} {\bibinfo {title} {{Strong} nodeless pairing on separate electron
			{Fermi} surface sheets in {(Tl,K)Fe$_{1.78}$Se$_{2}$} probed by {ARPES}},\
	}\href {https://doi.org/10.1209/0295-5075/93/57001} {\bibfield  {journal}
		{\bibinfo  {journal} {EPL (Europhysics Letters)}\ }\textbf {\bibinfo {volume}
			{93}},\ \bibinfo {pages} {57001} (\bibinfo {year} {2011})}\BibitemShut
	{NoStop}%
	\bibitem [{\citenamefont {Zhang}\ \emph {et~al.}(2011)\citenamefont {Zhang},
		\citenamefont {Yang}, \citenamefont {Xu}, \citenamefont {Ye}, \citenamefont
		{Chen}, \citenamefont {He}, \citenamefont {Xu}, \citenamefont {Jiang},
		\citenamefont {Xie}, \citenamefont {Ying}, \citenamefont {Wang},
		\citenamefont {Chen}, \citenamefont {Hu}, \citenamefont {Matsunami},
		\citenamefont {Kimura},\ and\ \citenamefont
		{Feng}}]{NodelessSuperconductingGapzhang2011}%
	\BibitemOpen
	\bibfield  {author} {\bibinfo {author} {\bibfnamefont {Y.}~\bibnamefont
			{Zhang}}, \bibinfo {author} {\bibfnamefont {L.~X.}\ \bibnamefont {Yang}},
		\bibinfo {author} {\bibfnamefont {M.}~\bibnamefont {Xu}}, \bibinfo {author}
		{\bibfnamefont {Z.~R.}\ \bibnamefont {Ye}}, \bibinfo {author} {\bibfnamefont
			{F.}~\bibnamefont {Chen}}, \bibinfo {author} {\bibfnamefont {C.}~\bibnamefont
			{He}}, \bibinfo {author} {\bibfnamefont {H.~C.}\ \bibnamefont {Xu}}, \bibinfo
		{author} {\bibfnamefont {J.}~\bibnamefont {Jiang}}, \bibinfo {author}
		{\bibfnamefont {B.~P.}\ \bibnamefont {Xie}}, \bibinfo {author} {\bibfnamefont
			{J.~J.}\ \bibnamefont {Ying}}, \bibinfo {author} {\bibfnamefont {X.~F.}\
			\bibnamefont {Wang}}, \bibinfo {author} {\bibfnamefont {X.~H.}\ \bibnamefont
			{Chen}}, \bibinfo {author} {\bibfnamefont {J.~P.}\ \bibnamefont {Hu}},
		\bibinfo {author} {\bibfnamefont {M.}~\bibnamefont {Matsunami}}, \bibinfo
		{author} {\bibfnamefont {S.}~\bibnamefont {Kimura}},\ and\ \bibinfo {author}
		{\bibfnamefont {D.~L.}\ \bibnamefont {Feng}},\ }\bibfield  {title} {\bibinfo
		{title} {{Nodeless Superconducting Gap in {{A$_x$Fe$_2$Se$_2$}}
				({{A}}={{K}},{{Cs}}) Revealed by Angle-Resolved Photoemission
				Spectroscopy}},\ }\href {https://doi.org/10.1038/nmat2981} {\bibfield
		{journal} {\bibinfo  {journal} {Nature Materials}\ }\textbf {\bibinfo
			{volume} {10}},\ \bibinfo {pages} {273} (\bibinfo {year} {2011})}\BibitemShut
	{NoStop}%
	\bibitem [{\citenamefont {Dong}\ \emph {et~al.}(2015)\citenamefont {Dong},
		\citenamefont {Jin}, \citenamefont {Yuan}, \citenamefont {Zhou},
		\citenamefont {Yuan}, \citenamefont {Huang}, \citenamefont {Hua},
		\citenamefont {Sun}, \citenamefont {Zheng}, \citenamefont {Hu}, \citenamefont
		{Mao}, \citenamefont {Ma}, \citenamefont {Zhang}, \citenamefont {Zhou},\ and\
		\citenamefont {Zhao}}]{PhysRevB.92.064515}%
	\BibitemOpen
	\bibfield  {author} {\bibinfo {author} {\bibfnamefont {X.}~\bibnamefont
			{Dong}}, \bibinfo {author} {\bibfnamefont {K.}~\bibnamefont {Jin}}, \bibinfo
		{author} {\bibfnamefont {D.}~\bibnamefont {Yuan}}, \bibinfo {author}
		{\bibfnamefont {H.}~\bibnamefont {Zhou}}, \bibinfo {author} {\bibfnamefont
			{J.}~\bibnamefont {Yuan}}, \bibinfo {author} {\bibfnamefont {Y.}~\bibnamefont
			{Huang}}, \bibinfo {author} {\bibfnamefont {W.}~\bibnamefont {Hua}}, \bibinfo
		{author} {\bibfnamefont {J.}~\bibnamefont {Sun}}, \bibinfo {author}
		{\bibfnamefont {P.}~\bibnamefont {Zheng}}, \bibinfo {author} {\bibfnamefont
			{W.}~\bibnamefont {Hu}}, \bibinfo {author} {\bibfnamefont {Y.}~\bibnamefont
			{Mao}}, \bibinfo {author} {\bibfnamefont {M.}~\bibnamefont {Ma}}, \bibinfo
		{author} {\bibfnamefont {G.}~\bibnamefont {Zhang}}, \bibinfo {author}
		{\bibfnamefont {F.}~\bibnamefont {Zhou}},\ and\ \bibinfo {author}
		{\bibfnamefont {Z.}~\bibnamefont {Zhao}},\ }\bibfield  {title} {\bibinfo
		{title} {({{Li}}$_{0.84}${{Fe}}$_{0.16}$){{OHFe}}$_{0.98}${Se}
			superconductor: {Ion}-exchange synthesis of large single-crystal and highly
			two-dimensional electron properties},\ }\href
	{https://doi.org/10.1103/PhysRevB.92.064515} {\bibfield  {journal} {\bibinfo
			{journal} {Phys. Rev. B}\ }\textbf {\bibinfo {volume} {92}},\ \bibinfo
		{pages} {064515} (\bibinfo {year} {2015})}\BibitemShut {NoStop}%
	\bibitem [{\citenamefont {Davies}\ \emph {et~al.}(2016)\citenamefont {Davies},
		\citenamefont {Rahn}, \citenamefont {Walker}, \citenamefont {Ewings},
		\citenamefont {Woodruff}, \citenamefont {Clarke},\ and\ \citenamefont
		{Boothroyd}}]{Davies2016}%
	\BibitemOpen
	\bibfield  {author} {\bibinfo {author} {\bibfnamefont {N.~R.}\ \bibnamefont
			{Davies}}, \bibinfo {author} {\bibfnamefont {M.~C.}\ \bibnamefont {Rahn}},
		\bibinfo {author} {\bibfnamefont {H.~C.}\ \bibnamefont {Walker}}, \bibinfo
		{author} {\bibfnamefont {R.~A.}\ \bibnamefont {Ewings}}, \bibinfo {author}
		{\bibfnamefont {D.~N.}\ \bibnamefont {Woodruff}}, \bibinfo {author}
		{\bibfnamefont {S.~J.}\ \bibnamefont {Clarke}},\ and\ \bibinfo {author}
		{\bibfnamefont {A.~T.}\ \bibnamefont {Boothroyd}},\ }\bibfield  {title}
	{\bibinfo {title} {Spin resonance in the superconducting state of
			{${\mathbf{Li}}_{1\ensuremath{-}x}{\mathbf{Fe}}_{x}{\mathbf{ODFe}}_{1\ensuremath{-}y}\mathbf{Se}$}
			observed by neutron spectroscopy},\ }\href
	{https://doi.org/10.1103/PhysRevB.94.144503} {\bibfield  {journal} {\bibinfo
			{journal} {Phys. Rev. B}\ }\textbf {\bibinfo {volume} {94}},\ \bibinfo
		{pages} {144503} (\bibinfo {year} {2016})}\BibitemShut {NoStop}%
	\bibitem [{\citenamefont {Pan}\ \emph {et~al.}(2017)\citenamefont {Pan},
		\citenamefont {Shen}, \citenamefont {Hu}, \citenamefont {Feng}, \citenamefont
		{Park}, \citenamefont {Christianson}, \citenamefont {Wang}, \citenamefont
		{Hao}, \citenamefont {Wo}, \citenamefont {Yin}, \citenamefont {Maier},\ and\
		\citenamefont {Zhao}}]{Pan2017}%
	\BibitemOpen
	\bibfield  {author} {\bibinfo {author} {\bibfnamefont {B.}~\bibnamefont
			{Pan}}, \bibinfo {author} {\bibfnamefont {Y.}~\bibnamefont {Shen}}, \bibinfo
		{author} {\bibfnamefont {D.}~\bibnamefont {Hu}}, \bibinfo {author}
		{\bibfnamefont {Y.}~\bibnamefont {Feng}}, \bibinfo {author} {\bibfnamefont
			{J.~T.}\ \bibnamefont {Park}}, \bibinfo {author} {\bibfnamefont {A.~D.}\
			\bibnamefont {Christianson}}, \bibinfo {author} {\bibfnamefont
			{Q.}~\bibnamefont {Wang}}, \bibinfo {author} {\bibfnamefont {Y.}~\bibnamefont
			{Hao}}, \bibinfo {author} {\bibfnamefont {H.}~\bibnamefont {Wo}}, \bibinfo
		{author} {\bibfnamefont {Z.}~\bibnamefont {Yin}}, \bibinfo {author}
		{\bibfnamefont {T.~A.}\ \bibnamefont {Maier}},\ and\ \bibinfo {author}
		{\bibfnamefont {J.}~\bibnamefont {Zhao}},\ }\bibfield  {title} {\bibinfo
		{title} {Structure of spin excitations in heavily electron-doped
			{Li$_{0.8}$Fe$_{0.2}$ODFeSe} superconductors},\ }\href
	{https://doi.org/10.1038/s41467-017-00162-x} {\bibfield  {journal} {\bibinfo
			{journal} {Nature Communications}\ }\textbf {\bibinfo {volume} {8}},\
		\bibinfo {pages} {123} (\bibinfo {year} {2017})}\BibitemShut {NoStop}%
\end{thebibliography}

\end{document}